\documentclass[twocolumn,aps,nobalancelastpage,aps,raggedbottom]{revtex4-2}
\usepackage{times,soul}

\usepackage{amsmath,amsfonts,amssymb,graphicx,hyperref,epstopdf,xcolor,tikz,scalerel,natbib}
\usepackage{mathptmx,textcomp,braket}
\usepackage[T1]{fontenc}
\usepackage[utf8]{inputenc}
\usepackage{txfonts}
\usepackage{graphicx}
\usepackage{color}
\usepackage{graphicx}

\definecolor{lime}{HTML}{A6CE39}
\DeclareRobustCommand{\orcidicon}
{
	\begin{tikzpicture} 
	\draw[lime, fill=lime] (0,0) circle [radius=0.15] node[white] {{\fontfamily{qag}\selectfont \tiny ID}};
	\draw[white, fill=white] (-0.0625,0.095) 	circle [radius=0.007];
	\end{tikzpicture}
	\hspace{-2.2mm}
}
\newcommand\orcidID[1]{\href{https://orcid.org/#1}{\orcidicon}}

\newcommand{\beqa}{\begin {eqnarray}}
\newcommand{\eeqa}{\end {eqnarray}}

\hypersetup{colorlinks,citecolor=blue,filecolor=black,linkcolor=blue,urlcolor=blue}

\newcommand{\F}{\mathcal{F}}

\date{}
\def\be{\begin{equation}}
\def\ee{\end{equation}}
\def\v{\mathcal{V}}
\def\u{{\mathbf{U}}}

\def\A{\mathcal {A}}
\def\B{\mathcal {B}}
\def\N{\mathcal {N}}

\def\kk{\mathbf{k}}
\def\kperp{\mathbf{k}_{\perp}}  
\def\kpar{k_{\parallel}}
\def \nh{H~{\sc i~}}

\def\n{{\mathbf{\hat{n}}}}

\def\A{\mathcal{A}}
\def\vtheta{\vec {\mathbf{\theta}}}
\def\vu{\vec{\bf{U}}}

\begin{document}

\title{Intensity mapping of post-reionization 21-cm signal and  its cross-correlations as a probe of $f(R)$ gravity }

\author{Chandrachud B.V. Dash}\email[E-mail: ]{cb.vaswar@gmail.com}

\author{Tapomoy Guha Sarkar}
\email[E-mail: ]{tapomoy1@gmail.com}

\affiliation{Department of Physics, Birla Institute of Technology and Science - Pilani, Rajasthan, India}

\author{Anjan Kumar Sarkar}\email[E-mail: ]{anjansarkarbhp@gmail.com}

\affiliation{Raman Research Institute, Bangalore, India}

\date{\today}

\begin{abstract}
We propose the intensity mapping of the redshifted \nh 21-cm signal from the post-reionization epoch as a cosmological probe of $f(R)$ gravity.
We consider the Hu-Sawicki family of $f(R)$ gravity models characterized by a single parameter $f_{,R0}$. The $f(R)$ modification to gravity 
affects the post-reionization 21-cm power spectrum through the change in the growth rate of density fluctuations. We find that a radio interferometric observation with a SKA1-Mid like radio telescope in both auto-correlation and cross-correlation with galaxy weak-lensing and Lyman-$\alpha$ forest may distinguish $f(R)$ models from $\Lambda$CDM cosmology at a precision  which is competitive with other probes of $f(R)$ gravity.
\end{abstract}

\maketitle

\section{Introduction}

Einstein's general theory of relativity (GR) has
endured a complete century of intensive scrutiny, and has emerged as
the most successful theory of gravitation. Several tests on solar
system scales have proved its consistency on small scales \cite{De_Marchi_2020}.
However, modifications to the theory of gravity have often been
proposed as a way to explain the observed cosmic acceleration
\cite{Faraoni:2009myt}. Several observational evidences like Galaxy redshift surveys, Cosmic
Microwave Background Radiation (CMBR) observations  and supernovae surveys strongly
indicate that the energy budget of our universe is dominated by dark
energy- a fluid with energy-momentum tensor that
violates the strong energy condition \cite{Perlmutter_1997, Spergel_2003, Hinshaw_2003, Riess_2016}.  The cosmological
constant ($\Lambda$) treated as a fluid with an equation of state $p=
-\rho$ is the most popular candidate for dark energy in the framework
of classical general relativity \cite{Padmanabhan_2003}. However, the
$\Lambda $CDM, model suffers from several theoretical and
observational difficulties \cite{Riess_2016, Zhao_2017, weinberg1989cosmological, Carroll_2001}.  In the
matter sector, scalar fields have often been used to model various properties
of dynamical and clustering dark energy \cite{Ratra-Peebles_1988, Turner-White_1997, Steinhardt_1998, Picon-Mukhanov_2001, Bento-Sen_2002, TRC-paddy_2002, Paddy-Bagla_2003, Sami_2007}. Extensive literature
is available on the diversity of such models and their general
treatment using model-independent parametrizations \cite{CHEVALLIER_2001, PhysRevLett.90.091301, Barboza_2008}.

 Alternatively, a modification of Einstein's theory can mimic dark energy without requiring an exotic fluid \cite{amendola_tsujikawa_2010}.
 In  $f(R)$ theory, the Ricci scalar $R$ appearing in the  Einstein-Hilbert action, is  replaced by a general function of $R$   \cite{nojiri2007introduction, sotiriou2007metric,capozziello2008extended, sotiriou2010f} as
\begin{equation}
S = \frac{1}{2\kappa}\int d^4 x \sqrt{-g}f(R) + S_m
\end{equation}
where $\kappa = \frac{8\pi G_N}{c^4}$ and $S_m$ is the action for matter.
The $f(R)$ modification naturally  has its imprint on the background comsological evolution and growth of structures.

Tomographic intensity mapping of the neutral hydrogen \nh distribution \cite{Bull_2015, Obuljen_2018, param2}
using radio observations of the redshifted 21-cm radiation is a
powerful probe of cosmic evolution and structure formation in the post
reionization epoch \cite{poreion0, poreion1, poreion2, poreion3, poreion4, poreion5, poreion6, poreion7, poreion8}.  The epoch of reionization is believed to be  completed
by redshift $z \sim 6$ \cite{Gallerani_2006}. Following the complex phase transition
characterizing the epoch of reionization (EoR), some remnant neutral
hydrogen remained clumped in the dense self shielded Damped Ly-$\alpha$ (DLA)
systems \cite{wolfe05}.  These DLA systems are the dominant source of the \nh 21-cm
signal in the post-reionization era. Intensity mapping experiments aim
to map out the collective \nh 21-cm radiation without resolving the
individual gas clouds.  The redshifted 21-cm signal from
the post-reionization epoch as a biased tracer \cite{Sarkar_2017, Sarkar_2016, Guha_Sarkar_2012, Bagla_2010}
of the dark matter distribution imprints a host of astrophysical and
cosmological information. It is, thereby a direct probe of large scale
matter distribution, growth of perturbations and the expansion history
of the Universe. Observationally the post-reionization signal has two
key advantages - Firstly, since the Galactic synchrotron foreground scales
as $\sim {(1 + z)}^{2.6}$, the lower redshifts are far less affected by
the galactic foreground. Secondly, in the redshift range $ z \leq 6$
the astrophysical processes of the EoR are absent whereby the
background UV radiation field does not have any feature imprinted on
the 21-cm signal.

The $f(R)$ modification to gravity will affect the 21-cm power
spectrum through its signature on cosmic distances, the Hubble
parameter and the growth rate of density perturbations.  We consider a
Hu-Sawicki form of $f(R)$, and
investigate the possibility of differentiating such a modification
from the standard $\Lambda $CDM model. 

In this paper our objective is to make error projections on parameters of a $f(R)$ gravity theory using the post-reionization 21-cm power spectrum in  auto and cross-correlations.
For cross-correlation of the 21-cm signal we have considered two dark matter tracers: (a) galaxy weak lensing and (b) the Lyman-$\alpha$ forest.

We investigate observational strategies with the upcoming SKA  towards constraining $f(R)$
theories at precesion levels competitive if not significantly better
than the next generation of supernova Ia observations, galaxy surveys,
and CMB experiments.

 \section{Cosmology with $f(R)$ gravity } We consider a spatially flat
Universe comprising of radiation (density $\rho_{\gamma} $) and
non-relativistic matter (density $\rho_{m} $). In a $f(R)$ gravity
theory, the Einstein's field equation and its trace for a
Friedman-Lemitre-Robertson-Walker metric (FLRW) with a scale factor
$a(t)$ and Hubble parameter $H = \frac{\dot{a}(t)}{a(t)}$ reduces to
\cite{tsujikawa2009dispersion}
\begin{eqnarray}
3 H^2 f_{,R} - \frac{1}{2}\left ( R f_{,R}- f \right )  + 3H \dot{f}_{,R} = \kappa^2 (\rho_{m} + \rho_{\gamma}) \\
H   f_{,R} -2 f_{,R}\dot{H}- \ddot{f}_{,R } = \kappa^2\left(  \rho_m + \frac{4}{3} \rho_{\gamma} \right)
\end{eqnarray}
where $ f_{,R} = \partial f(R) /\partial R$ and the $``." $ denotes a
differentiation with respect to the cosmic time $t$.  The Ricci scalar
$R$ is given by $R = 6( 2H^2 + \dot H) $.  It is convenient to express 
the above equations in terms of the following set of dimensionless
variables $x_{1} \equiv -\frac{\dot{f}_{,R}}{Hf_{,R}} $, $ x_{2}
\equiv - \frac{f}{6H^2f_{,R}}$, $x_{3} \equiv \frac{R}{6H^2}$, $x_{4}
\equiv \frac{\kappa^2 \rho_{\gamma}}{3H^2f_{,R}}$.  In terms of these
quantities the dynamical evolution of the density parameters can
obtained by solving the following set of autonomous first order
differential equations \cite{tsujikawa2009dispersion}
\begin{eqnarray}
x_1' = -1-x_3 -3x_2 + x_1^2 - x_1 x_3 + x_4 \\
x_2' = \frac{x_1x_3}{m} -x_2(2x_3 - 4 - x_1) \\
x_3' = -\frac{x_1 x_3}{m} - 2x_3(x_3-2) \\
x_4' = -2x_3x_4 + x_1x_4
\end{eqnarray}
where $ ' = d/d \ln(a)$ and $m$ measures the deviation from $\Lambda$CDM model defined as  
$m \equiv \frac{d \ln f_{,R}}{d \ln R} = \frac{Rf_{,RR}}{f_{,R}}$.
These equations form a 4-dimensional coupled dynamical system which can be integrated numerically for a given  $f(R)$ and with suitable initial conditions. 
 Solution to  the above coupled ODEs can be used to determine the dynamics of the  density parameters and map a $f(R)$ gravity theory to a dark energy with an effective equation of state (EoS) $w_{eff}(z)$  as 
\begin{eqnarray}
\Omega_{m} \equiv \frac{\kappa^2\rho_{m}}{3H^2f_{,R}} = 1-(x_1+x_2+x_3+x_4) \\
\Omega_{\gamma} \equiv x_{4} \\
\Omega_{_{DE}} \equiv x_1 + x_2 + x_3 \\
w_{eff} \equiv -\frac{1}{3}(2x_3 -1)
\end{eqnarray}

A wide variety of $f(R)$ models have been proposed \cite{Hu_2007, carroll2004cosmic, nojiri2007introduction, capozziello2002curvature}.  The
functional form of $f(R)$ is chosen so that the model is
phenomenologically satisfactory.  We expect  the $f(R)$ cosmology
to be   indistinguishable from the $\Lambda $CDM at high redshifts where
the latter is  well constrained from CMBR observations. At low redshifts the
accelerated expansion history should be close to the $\Lambda $CDM
predictions and on Solar system scales the proposed $f(R)$ model should be consistent with the $\Lambda $CDM model as a
limiting case.

We consider the $f(R)$ gravity model proposed by Hu-Sawicki (HS),
where the functional form of $f(R)$ is given by
\cite{Hu_2007, amendola_tsujikawa_2010}
\begin{equation}
\label{eqn:frmodel}
f(R) = R - \mu R_c \frac{(R/R_c)^{2}}{(R/R_c)^{2}+1}
\end{equation}
Here $\mu$ and $R_c$ are two non-negative parameters in the model
where $R_c$ is the present day value of the Ricci scalar. The
expansion rate $H$ for a viable $f(R)$ gravity theories is expected to
be close to the concordance $\Lambda$CDM \cite{hu2007parametrized}
predictions. The quantity $f_{,R}$ plays a crucial role to quantify
the deviation of $f(R)$ gravity models from GR whereby $f_{,R}$
behaves like an extra degrees of freedom that acts similar to a scalar
field. We may write \be  f_{,R} = -2f_0 \frac{R}{H_0^2}\left[ 1+ \left(
  \frac{R}{R_c} \right)^2 \right]^{-2} \ee
  with  $ |f_0| \equiv (\mu H_0^2)/R_c$ as the only free parameter.

To recover standard GR results in Solar system tests, the present day value of $f_{,R}$ is restricted to $ \log_{10}|f_{,R0}| < {-6}$
\cite{Hu_2007}. Further,  the second
derivative $f_{,RR} = d^2f(R)/dR^2 > 0$ in order to avoid ghost and
tachyonic solutions \cite{Amendola_2007}. Weak Lensing peak abundance
studies have provided strong constraints on $\log_{10} |f_{,R 0}|<-
4.82$ and $<-5.16$ with WMAP9 and Planck15 priors, respectively.
Tight constraints are also obtained from weak lensing peak statistics
study with $\log_{10} |f_{,R 0}|<-4.73$ (WMAP9) and $\log_{10} |f_{,R
  0}|<-4.79$ (Planck2013) (\cite{Liu_2016}).  In our work we adopt the
fiducial value $\log_{10} |f_{,R 0}| =-5$ from observations \cite{cataneo2015new,
  Liu_2016}.

Growth of large scale structure (LSS) offers a unique possibility to
constrain cosmological models. The quantity of interest is the
growth rate of matter density perturbations $ f_g (k,z) \equiv
\frac{d~\ln~\delta_m (k,z)}{d~\ln~a} $ which is sensitive to the
expansion history of the Universe.  In the linear perturbation theory,
and on sub-horizon scales ($k/a$ $>>$ $H$) the evolution of matter
density perturbations $\delta_m(k, z)$ is dictated by the differential
equation
\cite{tsujikawa2009dispersion,boisseau2000reconstruction,song2007large,tsujikawa2008constraints}.
\begin{equation}
\label{eqn:growthfr}
\ddot{\delta}_m + 2H\dot{\delta}_m - 4\pi G_{eff}(a,k)\rho_m \delta_m \simeq 0
\end{equation}
where $G_{eff}$ is an effective gravitational constant which is related to standard Newtonian gravitational constant ($G_N$) as 
\begin{equation}
\label{eqn:growth}
G_{eff}(a,k) = \frac{G_N}{f_{,R}}\left [ 1 + \frac{(k^2/a^2)(f_{,RR}/f_{,R})}{1+3(k^2/a^2)(f_{,RR}/f_{,R})} \right ] 
\end{equation}
In $f(R)$ theories $G_{eff}$ is a scale dependent function \cite{baghram2010structure}. The scale dependence of the growing mode of density fluctuations is widely exploited to differentiate the structure formation  beyond standard model of cosmology. In obtaining the approximate  equation (\ref{eqn:growthfr}) we have incorporated the assumption that oscillating modes are negligible compared to the modes induced by matter perturbations and also $\dot{f_{,R}} \approx 0$ on sub-horizon scales of interest \cite{tsujikawa2009dispersion}. 

\begin{figure}[h]
\begin{center}
\includegraphics[width=9.5cm]{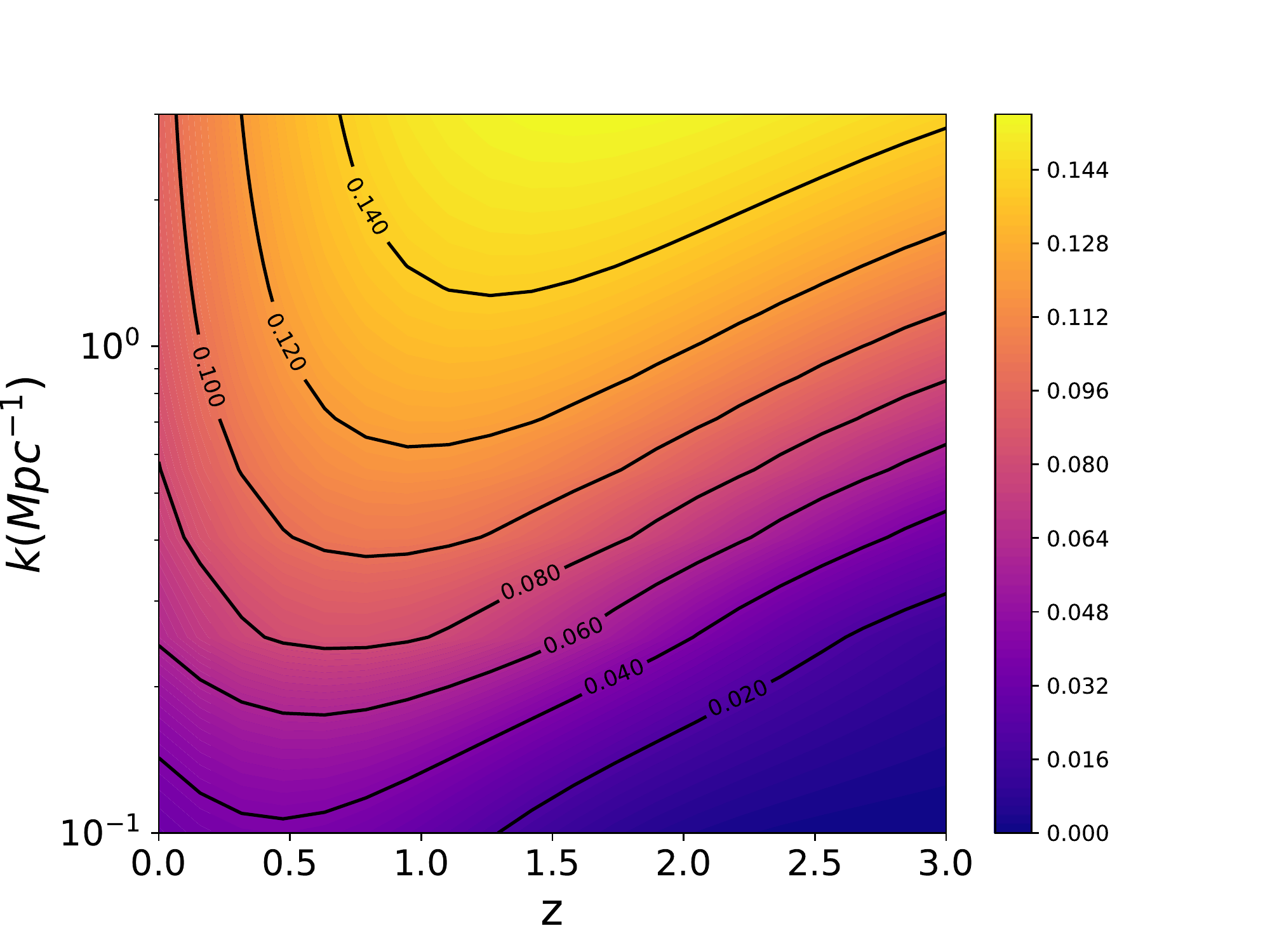}
\caption{The  figure shows the departure of the growth rate $f_g(z, k)$ for the $f(R)$ theory with $\log_{10} |f_{,R 0}| = - 5 $ from the $\Lambda $CDM prediction. We plot the  quantity $ | \frac{f_g - f_g^{{\Lambda }CDM}} {f_g^{{\Lambda}CDM}}|$ in the $(z, k)$ plane.}
\label{fig:fRcosmo}
\end{center}
\end{figure}

Figure (\ref{fig:fRcosmo}) shows the departure of the  the growth rate $f_g(z, k)$ for the $f(R)$ theory with $\log_{10} |f_{,R 0}| = - 5 $ from the $\Lambda $CDM prediction.
We know that the growth rate is scale independent and depends only on redshift for the $\Lambda $CDM model. Thus the $k-$dependence seen in the figure arises purely from the $f(R)$ modification to gravity.
Since different modes grow differently, the evolution has an additional contribution towards changing the shape of the cosmological power spectrum. 
The departure is small at very low redshifts and also very high redshifts and increases monotonically with $k$ for a given redshift. 
We find that a departure of  $>12\%$  is seen in the redshift window $0.5 < z < 1.5$ for $ k > 0.5 Mpc^{-1}$.

\begin{figure}[h]
\begin{center}
\includegraphics[width=8.5cm]{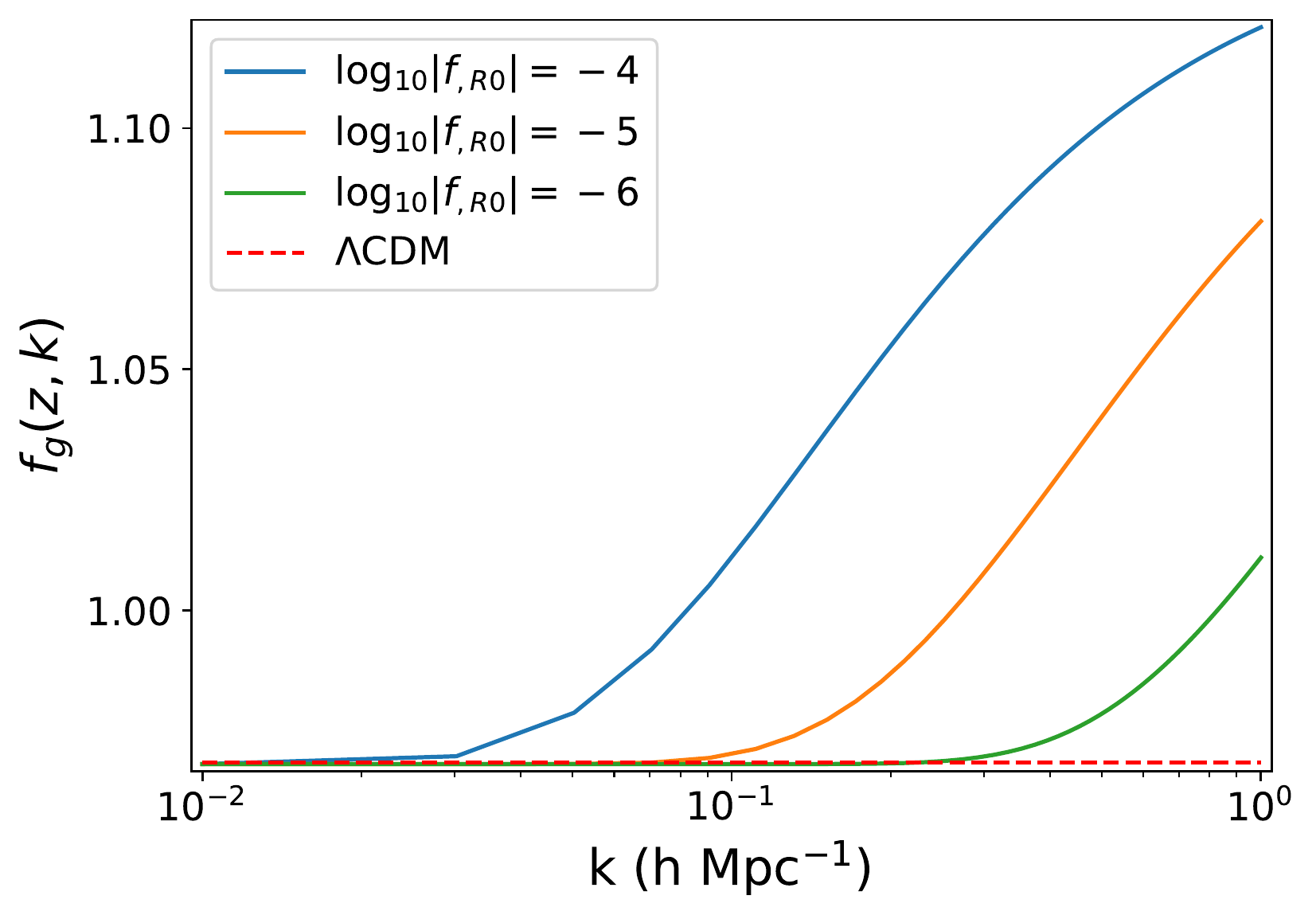}
\caption{The  figure shows the growth rate $f_g(z, k)$ for the $f(R)$ theory of $\log_{10} |f_{,R 0}| = - 5$ model. We also shown the results for $\log_{10} |f_{,R 0}| = - 4$ and $\log_{10} |f_{,R 0}| = - 6$ model for comparison purpose. The dotted bottom shows scale independent $\Lambda$CDM prediction. The fiducial redshift chosen to be $z = 2.3$.}
\label{fig:fRcosmo11}
\end{center}
\end{figure}

In Figure (\ref{fig:fRcosmo11}) we have shown the linear growth rate $f_g(z,k)$ for $\Lambda$CDM and $f(R)$ with $\log_{10} |f_{,R0}|=-5$ at a redshift $z=2.3$. At smaller scales (large $k$ modes) the scale dependent growth become more prominent and larger scales (small $k$ modes) $f(R)$ gravity coincide with $\Lambda$CDM. We have also shown the $f_{g}(k,z)$ for $\log_{10} |f_{,R 0}| = - 4$ and $\log_{10} |f_{,R 0}| = - 6$ gravity model for caparison purpose only. 

\subsection*{Matter power spectrum}

$f(R)$ gravity has a significant impact on structure formation in low density regions through a scale dependent growth factor because of enhancement of gravitational forces.
The modification to the force law in modified gravity theories is highly constrained from local tests \cite{will2014confrontation}. 
It is also well studied that $f(R)$ modification to gravity will induce non-linearities in the power spectrum through mechanisms  like chameleon \cite{mota2007evading}, dilaton  effect \cite{brax2014early} etc.
The shape of the matter power spectrum is sensitive to the  choice of cosmological model and as such it is sensitive probe of the underlying theory of gravity or dark
energy.
We model the power spectrum in $f(R)$ gravity models as 
\be
\label{eqn:supmps}
P_{f(R)} (k) =\frac{ P_{Lin}}{(1+k^2/k_{trunc}^2)^2} ~e^{-(k/k_s)^2}
\ee
where $P_{Lin}$ is the linear matter power spectrum. In our analysis we have used the analytic fitting function by Hu-Eisenstein for $P_{Lin}$ \cite{eisenstein1998baryonic}. We have used the fitting parameters $(k_{trunc}, k_s)$ for the suppressed matter power spectrum from \cite{brax2019lyman} for $f(R)$ gravity models.
$P(k,z)$ is obtained by multiplying the square of the growing mode with this. We remind ourselves that the growing mode is scale dependent for $f(R)$ models and is scale independent for $\Lambda$CDM model.

\begin{figure}[h]
\begin{center}
\includegraphics[width=8.5cm]{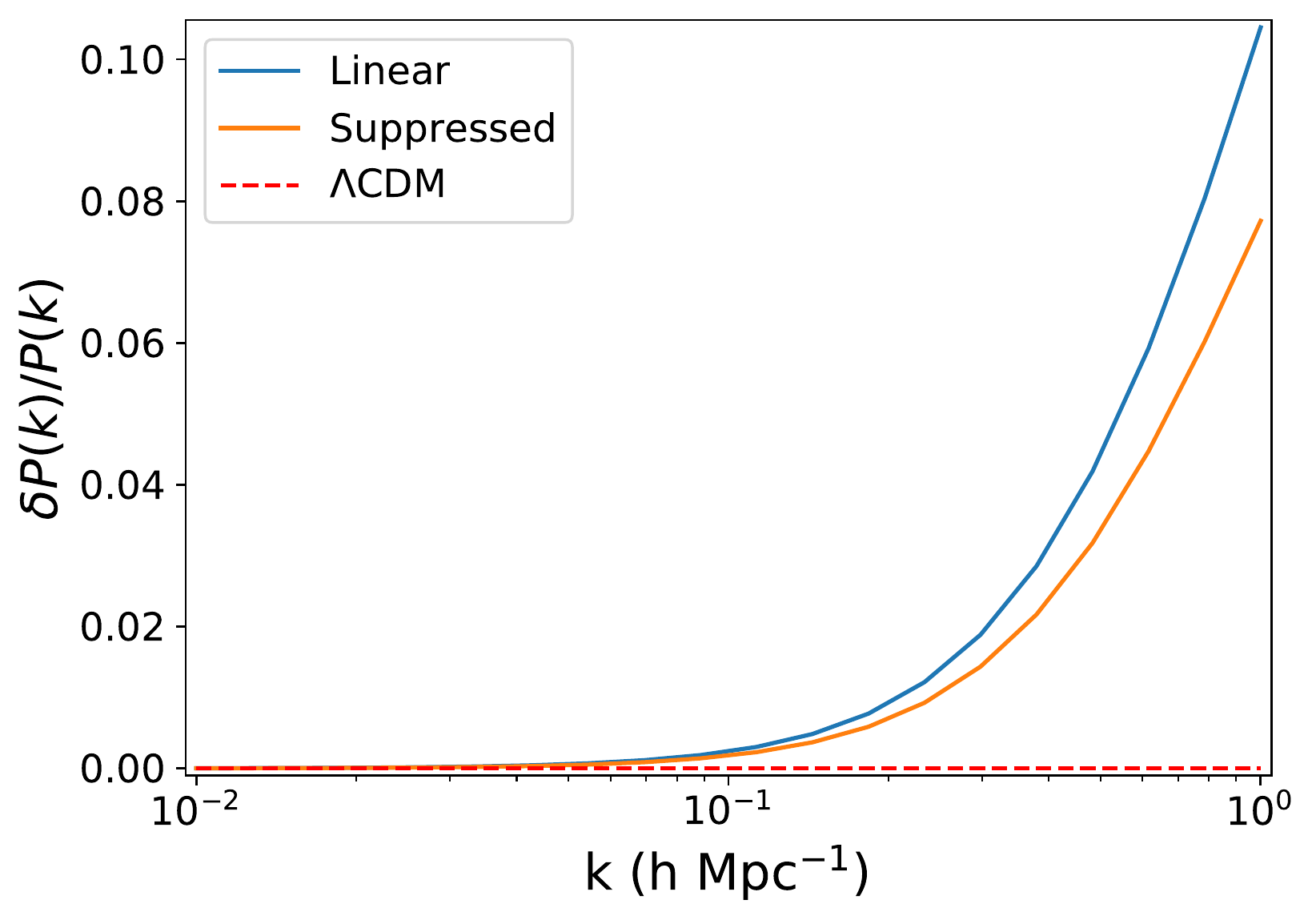}
\caption{The  figure shows the departure of the matter power spectrum $P(z, k)$ for the $f(R)$ theory with $\log_{10} |f_{,R 0}| = - 5 $ from the $\Lambda$CDM prediction. We plot the  quantity $  \frac{P_{f(R)}(k,z)  - P_{\Lambda CDM}(k,z) } {P_{\Lambda CDM}(k,z)}$ at redshift $z = 2.3$.}
\label{fig:fRmps}
\end{center}
\end{figure}

Figure (\ref{fig:fRmps}) shows the relative deviation of matter power spectrum of $f(R)$ gravity theory from $\Lambda$CDM at a fiducial redshift $z=2.3$. The topmost curve corresponds to  linear theory prediction and the one below shows the suppressed matter power spectrum due to the additional factor introduced in Eq(\ref{eqn:supmps}). The relative deviation of $P_{Lin}(k)$ from its $\Lambda$CDM counterpart grows at smaller scales, because the mass of the scalar field yields a characteristic scale dependence for the linear growing mode. 
Moreover, on  linear scales there is no additional chameleon screening mechanism.

Many simulation result shows that the deviation is significantly suppressed due to screening mechanism \cite{li2013non,arnold2019modified}. The additional prefactor in equation (\ref{eqn:supmps}) is fitted for mildly nonlinear behavior and reproduces the suppressed matter power spectrum with sub percent accuracy without requiring the full non linear simulations (refer Fig:5 in \cite{li2013non}). 

\section{The 21-cm signal from the post-reionization era}
Bulk of the low density hydrogen gets completely ionized
by the end of the reionization epoch  around $ z \sim 6$ \cite{Gallerani_2006}. 
A small fraction of \nh that survives the process of reionization is believed to remain confined in the over-dense regions of the
IGM. These clumped, dense damped Lyman-{$\alpha$} systems (DLAs) \cite{wolfe05} remain neutral as they are self shielded
from the background ionizing radiation. They store   $\sim 80\%$ of the \nh  at $z<4$ \cite{proch05} with \nh
column density greater than $ 2 \times 10^{20}$atoms/$\rm cm^2$
\cite{xhibar, xhibar1, xhibar2} and are the dominant source of the 21-cm radiation in the post-reionization epoch. The clustering properties
of these DLA clouds suggest that they are associated
with galaxies and located in regions of  highly non-linear matter over densities
\cite{coo06, zwaan, nagamine}. The 21-cm signal from the
post-reionization epoch has been extensively studied \cite{poreion9, poreion6,  poreion1, poreion2, poreion7, poreion8, poreion0}. 
The emitted flux from individual
clouds is extremely  weak ($< 10\mu \rm Jy$). These individual DLA clouds are
unlikely to be detected in radio observations, even with futuristic telescopes. However, in an intensity mapping experiment one does not aim to resolve the individual sources.
The collective emission  
forms a diffused background in all radio-observations at the observation
frequencies less than $1420$MHz. Fluctuations of this signal 
 on the sky plane and across redshift,  maps out the three dimensional
tomographic image of the Universe. 

Several  assumptions simplify the modeling of the
post-reionization \nh signal. These are either motivated from implicit
observations or from numerical simulations.

\begin{itemize}
\item
In the post-reionization epoch there is an enhancement of population of the triplet state of \nh 
due to the  Wouthheusen field coupling. This makes the spin temperature $T_s$ much greater than the CMB
temperature $T_{\gamma}$. Thus, the 21-cm
radiation is seen in emission in this
epoch against the background CMBR \cite{madau199721, bharad04, zaldaloeb}.
For $ z \leq 6$ the spin temperature and the  gas
kinetic temperature remains strongly coupled through Lyman-$\alpha$
scattering or collisional coupling \cite{madau97}.
 \item 
Extensive study of the Lyman-{$\alpha$} absorption lines in quasar spectra 
indicates that in the redshift range $ 1 \leq z \leq 3.5$ the
cosmological density parameter of the neutral gas has  a  value $ \Omega_{gas}
\sim 10^{-3}$ \cite{proch05}.
Thus the  mean neutral
fraction  is $\bar{x}_{\rm HI} = \Omega_{gas}/\Omega_b \sim 2.45\times 10^{-2}$.
which does not evolve in the entire redshift range $ z \leq 6$.
\item On the large cosmological  scales of interest, \nh peculiar velocities are assumed to be determined by the dark
matter distribution. Thus, peculiar velocity manifests as a  redshift space
distortion  anisotropy in the 21-cm power spectrum.
\item
  The discrete nature of DLA sources is not considered. The corresponding Poisson noise owing
  to this discrete sampling is neglected assuming that the number density of
  the DLA emitters is very large \cite{poreion8}.
\item \nh perturbations are generated by a Gaussian random
  process. We do not consider any non-gaussianity and thereby the statistical
  information is contained in the two-point correlation or the power spectrum.
\item  
Galaxy redshift surveys and numerical
simulations show that the galaxies are a biased tracers of  the underlying dark matter
distribution \cite{dekel, mo, yosh}. If we assume that \nh in the  post-reionization epoch 
is housed predominantly in dark matter haloes,  we may expect  the gas to trace the underlying dark
matter density field  with a bias $b_T(k, z)$ defined as $
   {b}_T(k, z) = {\left [\frac{P_{\rm HI}(k, z)}{P(k, z)}\right ]}^{1/2}
$where  ${P}_{\rm HI}(k, z)$ and ${P}(k, z)$ denote the \nh and
  dark matter power spectra respectively. 
  The bias function quantifies the nature of \nh clustering in the post-reionization
  epoch. Further, the fluctuations in the ionizing background
may also contribute to  $b_T(k,z)$ \cite{poreion0}.
 On  scales below  the Jean's length, the linear density contrast of \nh
gas is related to the dark matter density contrast though a scale
dependent function \cite{fang}. However, on large scales the bias is known to be
scale-independent, though the scales above which the
bias is linear, is sensitive to the redshift being probed.
Several authors have now demonstrated the nature of \nh bias using N-body simulations
\cite{Bagla_2010, Guha_Sarkar_2012, Sarkar_2016, Carucci_2017}.
The simulations are based on the principle of populating dark matter halos in a certain mass range with gas 
and thereby identifying them as DLAs. 
These  simulations show  that the large scale  linear bias grows
monotonically with redshift for $1< z< 4$ \cite {Mar_n_2010}. This feature is
shared by galaxy bias as well \cite{fry, mo, moo}.
There is a  steep rise of the 21-cm bias on small scales. This is because of the absence of small mass halos as is expected from the CDM power spectrum and consequently the \nh being distributed only in larger halos. A fitting formula for the bias $b_T(k, z)$ as a function of both redshift $z$ and
scale $k$  has been obtained from numerical simulations \cite{Guha_Sarkar_2012, Sarkar_2016} of the post-reionization signal as
\be
\label{eqn:bias}
b_{T}(k,z) = \sum_{m=0}^{4} \sum_{n=0}^{2} c(m,n) k^{m}z^{n}
\ee
The critical density for collapse is smaller in $f(R)$ gravity models which leads to a significant suppression of bias i.e $b^{f(R)}_{T}<b^{GR}_{T}$.
This has been seen in numerical simulations  \cite{aviles2018nonlinear,arnold2019modified}. Though the fitting function for  $b_{T}$  (\ref{eqn:bias}) is obtained from $\Lambda$CDM simuation, we used the same form for $f(R)$ gravity assuming the bias is not significantly different in the redshifts of our interest. However we have kept the  bias as free parameter which we have eventually marginalized over. We kept the third order component of polynomial bias as the free parameter because we know on large scale the bias is completely indistinguishable from $\Lambda$CDM and suppression shows up  only on small scales. For $\log_{10} |f_{,R 0}| = - 5$ model our used bias fitting function is well within the error bars and can be used safely \cite{aviles2018nonlinear}.

We have used these simulation results in our modeling of the post-reionization
epoch.
\end{itemize}
Adopting all the assumptions discussed above,  the power spectrum of post-reionization  \nh  21-cm brightness temperature fluctuations from redshift
$z$
is  given by (\cite{bharad04, param3})
\be
P_{\rm HI}({\bf{k}}, z)  = \bar{T}(z)^2 \bar{x}_{\rm HI}^2 b_T(k,z)^2 {( 1  + \beta_T (k,z) \mu^2)}^2  P(k,z)
\ee
where $\mu
={\bf{\hat{k}}}\cdot{\bf{\hat{n}}}$, $\beta_T(k, z) = f_g(k, z)/b_T(k, z)$, 
and 
\be
\bar{T}(z)=4.0 \, {\rm mK}\,\,(1+z)^2  \, \left(\frac{\Omega_{b0}
  h^2}{0.02}\right)  \left(\frac{0.7}{h} \right) \frac{H_0}{H(z)}
\ee
The term $ f_g(z, k)  \mu^2$ has its origin in the  \nh peculiar
velocities \cite{poreion2, bharad04} which, as we mentioned, is also sourced
by the dark matter fluctuations.
\begin{figure}[h]
\begin{center}
\includegraphics[width=9.5cm]{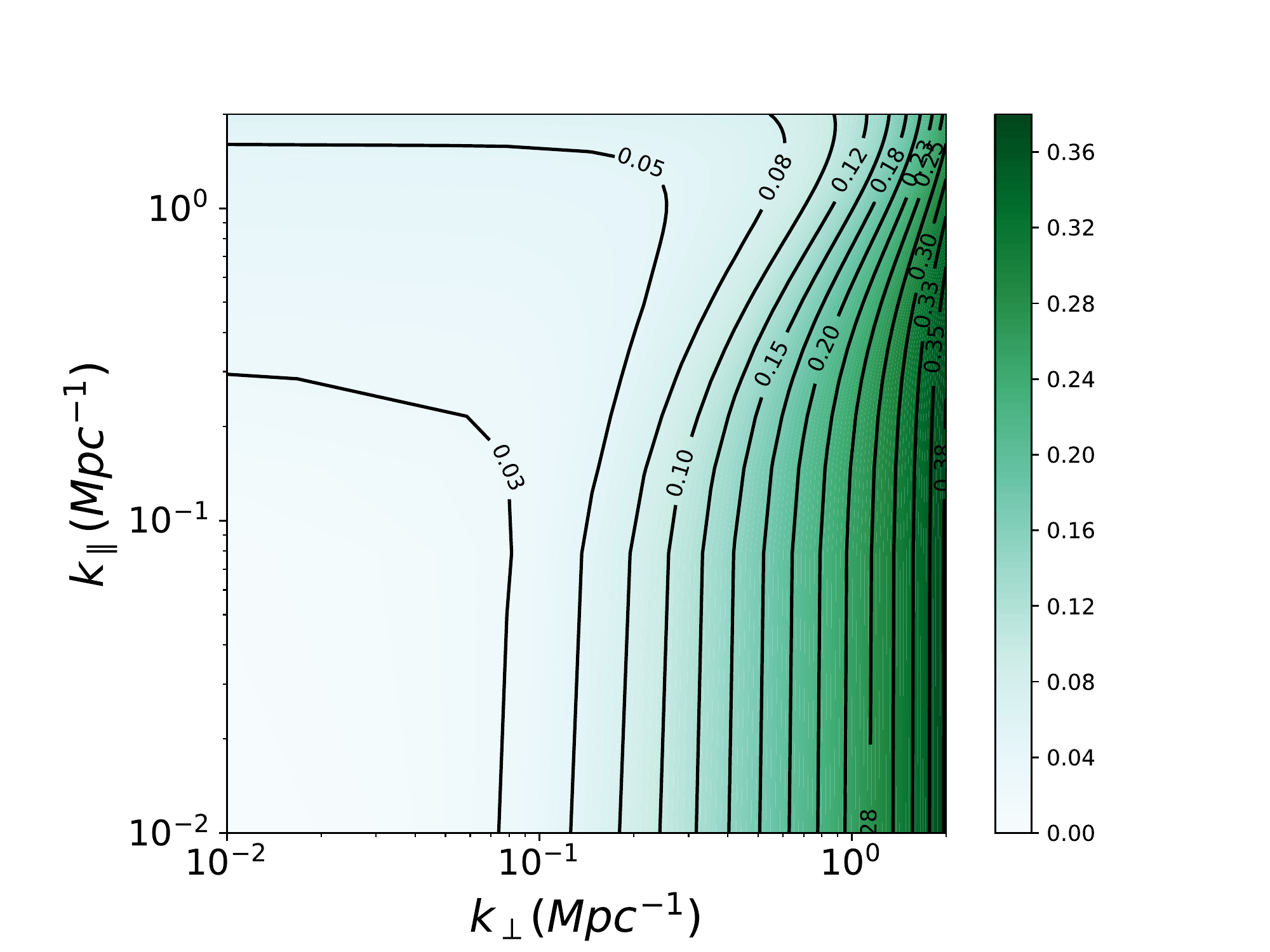}
\caption{The figure shows the 21-cm power spectrum in the $(\kpar, k_{\perp})$ space at the observing frequency $ \nu_0 = 710$MHz.}
\label{fig:signal} 
\end{center}
\end{figure}
 
The $f(R)$ modification affects the 21-cm power spectrum through the change in the redshift space distortion parameter $\beta_T(k,z)$, 
and $P(k,z)$. 
Figure (\ref{fig:signal}) shows the 21-cm power spectrum at $z=1$ in the $(\kpar, k_{\perp})$ space. The asymmetry in the signal is indicative of redshift space distortion and 
is sensitive to $\beta_T(k, z)$.
We emphasize that $({\bar{x}}_{HI}, \beta_T(k,z))$ along with the cosmological parameters completely model the post-reionization 21-cm signal.
We note that  the product $b_T^2 {\bar x}_{\rm HI}$ which
appears in the overall amplitude of the 21-cm signal is a largely unknown
parameter and depends largely on the \nh modeling. We shall, therefore  be interested in constraining the function $\beta_T(k,z)$ from some radio-interferometric observation of the signal. We shall marginalize our Fisher matrix projections over the overall amplitude to make error projections.

We shall now investigate the possibility of constraining the function $\beta_T(k, z)$ and thereby put observational bounds on $|f_{,R0}|$ from a radio-interferometric observation of the signal.

\subsection*{Observed 21-cm power spectrum}

The quantity of interest in  radio-interferometric observation is the complex visibility $\v(\u, \nu)$ measured as function of baseline $\u = (u, v) $ and observing frequency $\nu$.
Considering an observation frequency band width and defining $\Delta \nu$ as the difference from the central observing frequency, a  further Fourier transform in  $\Delta \nu$ gives us the visibility $v(\u, \tau)$ as a function of delay channel $\tau$. The measured visibility  can be written as a sum of signal $s(\u, \tau)$ and noise $ n(\u, \tau)$  as $v(\u, \tau) = s(\u, \tau) + n(\u, \tau) $.
The signal $s(\u, \tau)$ can be written as
\be s(\u_a, \tau_m ) = \frac{2k_B}{\lambda^2}    \int \frac{d^3 \kk }{(2 \pi)^3}  ~G( \kk,  \u_a, \tau_m)    ~\widetilde{\delta T_b}( \bf k)   \ee
where $ \widetilde{\delta T_b}(\bf k)$ denotes the fluctuations of the 21-cm brightness temperature in Fourier space. The transformation kernel $G$ is given by \[ G( \kperp, \kpar,  \u_a, \tau_m) =  \widetilde{\A}\left (\frac{\kperp  r}{2 \pi}  -  \u_a \right ) \widetilde{\B} \left ( \frac{\kpar r'}{2 \pi} -  \tau_m \right ) \]
where $\widetilde{\A}(\u)$ and $\widetilde{\B}(\tau)$ denote the Fourier transform of the telescope beam $A(\vec \theta)$ and the frequency response window function $B(\Delta \nu)$ respectively.
We use $r$ to  denote comoving distance to the observing redshift $z = (1420 ~{\rm MHz} / \nu)  - 1 $   and $r' = dr(\nu)/d\nu$.
The signal covariance matrix is defined as  \[  \langle s(\u_a, \tau_m )    s^*(\u_b, \tau_n ) \rangle = C^{S}_{ (a,m), (b,n)} \] and is given by  
\be
C^{S} = \left ( \frac{2k_B}{\lambda^2}  \right )^2 \frac{1}{r^2 r'} \int d^2 \u d \tau~ 
G ( \kk,  \u_a, \tau_m)
G^{*}( \kk,  \u_b, \tau_n) 
 \times P_{HI} ( \kk ) 
\ee
where $ \kk = ( \frac{2 \pi \u }{r}, \frac{2 \pi \tau }{r'})$. 
The noise in the visibilities measured at different baselines and frequency
channels are uncorrelated. If we define the noise covariance matrix as $C^N = \langle n ( \u_a,  \tau_m) ~n^*({\u_b, \tau_n}) \rangle $, we have 
 \be  C^{N}  =   \left (\frac { 2 k_B}{\lambda^2} \right ) ^2  \left (  \frac{ \lambda ^2 T_{sys} }{ A_e} \right ) ^2 \frac{B}{t}   \delta_{m,n} \delta_{a, b} \ee
 where $t$ is the correlator integration time and  $B$ is the observing bandwidth. The system temperature $T_{sys}$ can be written as a contribution from the instrument and the sky as  $ T_{sys} = T_{inst} + T_{sky}$, where
 $T_{sky} = 60 {\rm K}  \left (\frac{\nu }{300 ~{\rm MHz}} \right ) ^{-2.5}$.
 We first investigate the possibility of constraining the scale dependent function $\beta_T (k, z_{fid})$.
 We divide the observational range $k_{min}$ to $k_{max}$ into $N_{bin}$ bins and constrain the values of $\beta_T(k_i)$ at the middle of the bin $k_i$ using a Fisher matrix analysis.
The departure from the $\Lambda$CDM model for the fiducial  $\log_{10}|f_{,R 0}|<-5$ model for a range of $k$ values, peaks around $z \sim 1$. We choose the observational central frequency to be $710MHz$ corresponding to this redshift.
We first consider an  OWFA \cite{sarkar2018predictions, sarkar2017fisher, bharadwaj2015fisher} like array which is the upgraded version of the Ooty radio telescope and is expected to  operate as an linear 
radio-interferometric array.  The 
OWFA is a 530 m long and 30 m wide parabolic cylindrical reflector that is placed 
along the north-south direction on a hill that has the same slope ($\sim 11^{\circ}$)
as the latitude of the place. This makes it possible to track a given patch of 
sky by rotating the cylinder about the long axis of the telescope. The OWFA has 
1056 dipoles in total that are equally placed at $\sim 0.5 \, {\rm m}$ apart from 
each other along the long axis of the telescope. OWFA is capable of operating in
two independent simultaneous radio-interferometric modes - PI and PII. The OWFA PII
has 264 antennas in total, the radio signals from 4 consecutive dipoles have been 
combined to form a single antenna element. The OWFA PII has the smallest baseline
length, $d = 1.92 \, {\rm m}$ that corresponds to the distance between 
the two consecutive antennas in the array. The OWFA PII has an operating 
bandwidth, $B = 39 \, {\rm MHz}$ (for detailed specifications \cite{bharadwaj2015fisher}).
The full covariance matrix is given by 
\be C_{ab} =
C^S + \frac{ C^N}{N_r} \ee
where $N_r = 264 -a$ is the redundancy of the baselines. 
The Fisher matrix is given by 
\be 
F_{ij} = \frac{1}{2}\sum_m C^{-1}(m)_{ab} C(m)_{bc, i } C^{-1}(m)_{cd} C(m)_{bc, j } \ee
where $i$ and $j$ runs over the parameters $\beta_T(k_1), \beta_T(k_2), \dots \beta_T(k_{N_{bin}})$. 
The error on the $i^{th}$ parameter is  obtained from the Cramer Rao bound as $\sqrt {F^{-1}_{ii}}$. 
We find that in the $k-$range $ 0.06 < k < 1.32$ $\beta_T(k)$ can be measured in $4$ bins at $> 9\% $ for $500 \times 50 $hrs observation with  $50$ independent pointings.
Since the maximum departure of $\beta_T(k)$ from the $\Lambda $CDM is $\sim 11 \%$ in the $k-$range of interest, such an observation will at its best be able to distinguish between a $\log_{10}| f_{,R0}|  = -5$ at a $ \sim 1-\sigma$ level and $\log_{10}| f_{,R0}|= -4$ at $\sim 2-\sigma$ level. 
\begin{figure}[h]
\begin{center}
\includegraphics[width=8cm]{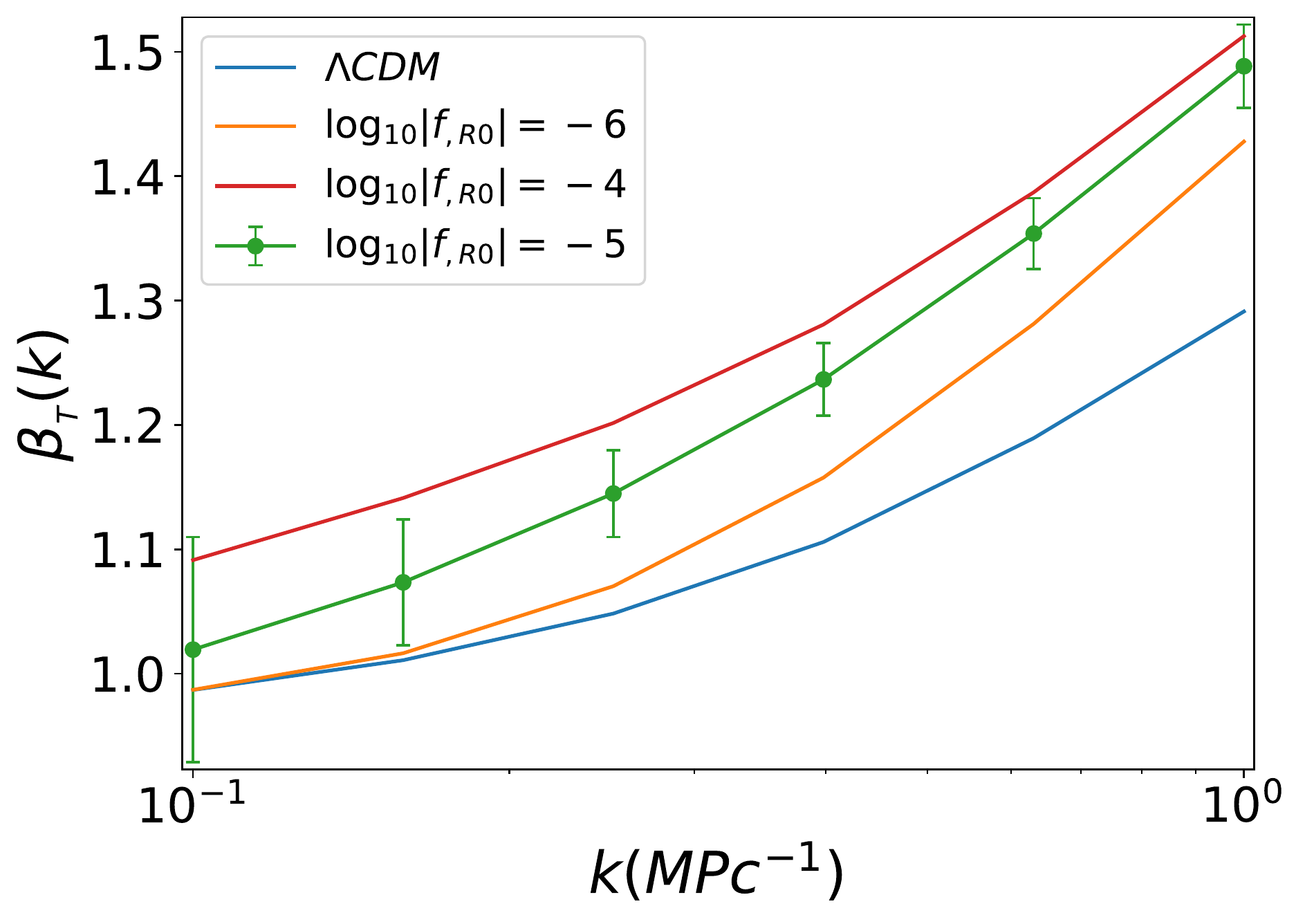}
\caption{ The figure shows the variation of $\beta_T(k, z_{fid})$ at the fiducial redshift $z_{fid} = 1$ for various Hu-Sawcki $f(R)$ models.
The $\Lambda $CDM prediction is also shown. We also show the $1-\sigma$ error bars on $\beta_T$ at $6$ logarithmically spaced $k-$ bins in the 
observed range of scales for the fiducial model with $\log_{10}|f_{,R0}| = -5$.}  
\label{fig:frbeta}
\end{center}
\end{figure}

For stronger constraints we now consider a { SKA1-mid } type of radio array.
We consider  a binning in visibility $\Delta \u$, and a total observing time $T_0$  
causing a  reduction of noise variance by a factor $N_p$ where $N_p$ is  the number of visibility pairs in the bin given by 
$N_p = N_{vis} ( N_{vis} - 1 ) /2  \approx  N_{vis}^2 /2 $
where $N_{vis}$ is the number of visibilities in the bin measured in time $T_0$. We may write 
\be N_{vis} = \frac{N_{ant} ( N_{ant} - 1) } {2} \frac{T_o}{t} \rho(\u ) \delta^2 U  \ee
where $N_{ant} $ is the total number of antennas in the array and $\rho(\u)$ is the baseline distribution function.
In general, the baseline distribution function is given by a convolution 
\be \rho (\u) = c \int d^2 {{\bf r}} \rho_{ant} ({\bf r}) \rho_{ant} ({\bf r} - \lambda \u ) \ee
Where $c$ is fixed by normalization of $\rho(\u)$ and $\rho_{ant}$ is the  distribution of antennas.
Further, if we assume a uniform frequency response over the entire observation  bandwidth $B$ and a Gaussian beam for the telescope 
\be
\int d\tau ~\widetilde{\B} ( \tau - \tau _m )  \widetilde{\B}^* ( \tau - \tau _n) = B \delta_{mn}, ~~{\rm and} \ee
\be \int d^2 \u  \A ( \u - \u_a )  \A^{*} ( \u - \u_b ) \approx \frac{\lambda^2}{ A_e }\delta_{a,b}
\ee
where, $A_e$ is the effective area of the antenna dishes. With these simplifications we may then write 
\be
C^S  \approx  \left ( \frac{2k_B}{\lambda^2}  \right )^2 \frac{ B \lambda^2 }{r^2 r' A_e} P_{HI} \left (\frac{2 \pi \u_a }{r}, \frac{2 \pi \tau_m }{r'} \right ) \delta_{m,n} \delta_{a, b} \nonumber
\ee

The 21-cm power spectrum is not spherically symmetric, due to redshift space distortion but is symmetric around the polar angle $\phi$. Using this symmetry, we would want to sum all the Fourier cells in
an annulus of constant $(k, ~ \mu = \cos \theta = \kpar / k )$  with radial width $\Delta k$  and
angular width $\Delta \theta$  for a statistical detection with improved SNR. The number
of independent cells in such an annulus is 
\be 
N_c  = 2 \pi k^2  \Delta k \Delta \mu \frac{Vol}{(2\pi)^3}  \ee
where the volume $Vol$ of the intensity mapping survey is given by $Vol = \frac{r^2 \lambda^2 r' B }{ A_e}$.
Thus, the  full covariance matrix may be written as 
\be
C^{Tot} = \frac{1}{\sqrt N_c} \left [ C^S + \frac {C^{N}}{N_p} \right ] \ee
The covariance matrix is diagonal owing to the binning in $U$ since different baselines which get correlated due to the telescope beam are now uncorrelated.
Further, to increase the sensitivity we consider the angle averaged power spectrum by averaging over $\mu$.
Thus we have
\be
P_{\rm HI}(k)  = \bar{T}(z)^2 \bar{x}_{\rm HI}^2 b_T^2 {\left ( 1  + \frac{2}{3}\beta_T + \frac{1}{5} \beta_T^2 \right )}  P(k,z)
\ee
and the corresponding variance is obtained by summing 
\be 
\delta P_{HI}(k) = \left [ \sum_{\mu} \frac{1}{\delta P_{HI}(k, \mu)^2} \right ] ^{-1/2} \ee
where $
\delta P_{HI}(k, \mu)  =  \frac{A_e r^2 r'}{\lambda^2 B} C^{Tot}$.

The fisher matrix for parameters $\lambda_i$  may be written as
\be
F_{ij} = \sum_{k}  \frac{1}{\delta P_{HI}^2( k)} \frac{\partial P_{HI}(k) }{\partial \lambda_i } \frac{\partial P_{HI}(k)}{\partial \lambda_j}\ee
We consider a 
radio telescope with an operational frequency range of $350 $MHz to  $14 $ GHz. We consider $250$  dish antennae each of diameter $15 m$ and efficiency $0.7$. To calculate the normalized baseline distribution function
We assume that baselines are distributed such that the antenna distribution falls off as $ 1/r^2$.  We also assume that there
is no baseline coverage below 30m. We assume $T_{sys} = 60K$ and an observation bandwidth of $128 MHz$. We assume $\Delta U = U_{min} = 50$ over which the signal is averaged.

Figure (\ref{fig:frbeta}) shows the variation of $\beta_T(k, z_{fid})$ at the fiducial redshift $z = 1$ corresponding to the observing central frequency of $710 MHz$. The monotonic rise of $\beta_T(k, z_{fid} = 1.0 )$ owes its origin to both the monotonic growth of $f_{g}(k)$ and also a slow decrease of $b_T(k, z_{fid}= 1.0)$ in the $k-$ range of interest. The behaviour is similar for different values of $\log_{10} | f_{,R 0} |$. The $\Lambda $CDM result is seen to coincide with the $f(R)$ prediction on large scales. We note that the $ \log_{10} | f_{,R 0} | = -6$ matches with the $\Lambda $CDM model for $ k < 0.15 Mpc^{-1}$. 
We consider a fiducial $ \log_{10} |f_{,R 0}| = -5 $ for our analysis. The $k-$range between the  smallest and largest baselines in binned 
as $ \Delta k = \alpha k$  where 
$ \alpha = \frac{1}{N_{bin}} \ln  (U_{max}/U_{min})$, with $(U_{min}, U_{max} ) = (50,~550)$. We consider $400 \times 50$ hrs observation in $50$ independent pointings.
 The $1-\sigma$ errors on $\beta_T(k_i)$ are obtained from the Fisher matrix analysis where the overall normalization of the power-spectrum is marginalized over.
 We find that for  $k > 0.4 Mpc^{-1}$, the $ \log_{10} | f_{,R 0} | = -5 $ can be differentiated from the $\Lambda $CDM model at a sensitivity of
 $ > 5 \sigma$ if we consider  $6$ $k-$bins. On larger scales $ k < 0.4 Mpc^{-1}$ the $f(R)$ models  with  $ -6 < \log_{10} | f_{,R 0} |<  -4$ remain statistically indistinguishable from the $\Lambda$CDM model. Thus, it appears that 21-cm observations of the post-reionization  epoch may only be able to constrain $f(R)$ theories on relatively small scales.
  
Instead of constraining the binned function $\beta_T(k)$, we investigate the possibility of putting bounds on $\log_{10} | f_{,R 0} |$ from the given observation. Marginalizing over the overall  amplitude of the power spectrum, we are thus interested in two parameters $(\Omega_{m0}, ~\log_{10} | f_{,R 0} |)$. The $1-\sigma$ bounds on $\log_{10} | f_{,R 0} |$ obtained from the marginalized Fisher matrix is given in the table below.
\begin{table} [h]
\begin{center}
\caption{The $68 \%$ ($1-\sigma $) marginalized errors on $\log_{10}| f_{,R0} |$  and $ \Omega_{m0} $}
\begin{tabular}{c|c| c} 
 \hline 
 \hline 
   Model &  $ \log_{10}| f_{R0} |$  & $ \Omega_{m0} $  \\ 
 \hline
 $f(R)$ & $ ~~~-5 \pm 0.62~~$  & $~~~  0.315 \pm 0.005~~~ $ \\
 \hline
\end{tabular}
\end{center}
\end{table}
 	Our error projection maybe compared with constraints obtained from other observational probes.
 	We find that our projected constraints are competitive with constraints obtained from diverse probes.
\begin{table} [h]
\label{tab:comparison}
\begin{center}
\caption{Bounds  on $p = \log_{10}| f_{,R0} |$  from other probes}
\begin{tabular}{c|c } 
 \hline 
 \hline 
   Probe of $f(R)$ gravity  & Bound on  $ \log_{10}| f_{,R0} |$   \\ 
 \hline
 GW Merger GW170817  & $ p < -2.52 $ \cite{PhysRevD.99.044056} \\
 Suyaev Zeldovich clusters PLANCK & $-5.81 < p <  -4.40$ \cite{2017PhRvD..95b3521P} \\
 Weak lensing Peak Statistics & $-5.16 < p <  -4.82$ \cite{Liu_2016} \\
 CMB + Cluster + SN + $H_0$ + BAO & $ p  < -3.89$ \cite{Fabian} \\
  \hline
  \hline
   \end{tabular}
    \end{center}
 	\end{table}
 	
 	The radio-interferometric observation of the post-reionization \nh 21-cm signal, thus holds the potential of providing 
robust constraints on $f(R)$ models.

\section{Cross-correlation of 21-cm signal with galaxy weak lensing }

Weak-lensing \cite{waerbeke2003, MUNSHI_2008} of background source galaxies by large scale structure (cosmic shear) has been extensively studied as a powerful cosmological probe \cite{Jain_1997, Hu_1999, Huterer_2002, Takada_2003, Heavens_2003, knox2004weak, Miyazaki_2007,  Hoekstra_2008, Takada_2009}.
The quantity of interest to us is the amplification matrix \cite{waerbeke2003, MUNSHI_2008} which quantifies the distortions due to gravitational lensing. These distortions allow to analyze large scale structures and map the matter distribution, on a broad range of scales.  Noting that scalar perturbations can not induce any rotation, one only has shear ($\gamma$)  and convergence ($\kappa$)  effects in the lensed distorted image of a galaxy. This weak shear/convergence signal is superposed on the intrinsic ellipticities and irregularities of background galaxy images \cite{treu2010strong}. We are interested in the statistical properties of these distortion fields.  The angular power spectrum of the shear field is identical to that of the convergence field whereby we shall only be looking at the convergence field.
 The 
Weak-lensing convergence field on the sky is given by a weighted line of sight integral \cite{waerbeke2003} of the overdensity field $\delta$
\be 
\kappa({\vec \theta}) =  \int_0^{\chi_s} ~ \A_{\kappa} (\chi)  \delta(\chi \vec \theta, \chi)    d\chi
\ee
where $\chi$ denotes the comoving distance and  
\be \A_{\kappa} (\chi) = \frac{3}{2 } \left(\frac{H_0}{c}\right)^2 \Omega_m{_{_0}}\frac{    g(\chi) ~\chi~} {a(\chi)}, ~~~~{\rm with}~~~~g(\chi) = \int_\chi^{\chi_s} n(z) \frac{dz}{d\chi'}   \frac{\chi^{\prime} - \chi}{\chi^{\prime}} d\chi^{\prime} 
\label{eq:lensingkernel} 
\ee
The weight function appearing in the kernel incorporates the all the  sources distributed according to a distribution function $n(\chi )$ upto $\chi_s$. 
We have assumed that the source galaxies are distributed as  \cite{Huterer_2006}
\be n(z) = n_0 \left (\frac{z}{z_0} \right )^\alpha exp \left [ - (z/z_0) \right ] ^{\beta}\ee
 In this work, we have considered a weak-lensing survey where $z_0 =  0.5$, $\alpha = 2$ and $\beta = 1$ \cite{Takada_2009}.
On small angular scales (typically for $\ell >10$)  where "flat sky " approximation is reasonable we can use
the Limber approximation \cite{Limber} in Fourier space and write the weak-lensing convergence angular power spectrum as

\begin{equation}
C_\kappa^{\ell} = \frac{9}{4} \left(\frac{H_0}{c}\right)^4 \Omega^2_m{_{_0}} \int_0^{\chi_s} \frac{g^2(\chi)}{a^2{(\chi)}} P\left(\frac{\ell}{\chi},\chi\right) d\chi
\end{equation}
where $P$ denotes the matter power spectrum. 
The noise for the convergence angular power spectrum is  given by $\Delta C_\kappa^{\ell} $  where 
\begin{equation}
\Delta C_\kappa^{\ell} = \sqrt{\frac{2}{(2{\ell}+1)f_{sky}}} \left( C_\kappa^{\ell} + \frac{\sigma_\epsilon^2}{n_g}\right)
\end{equation}
Here, the Poisson noise is dictated by the total galaxy count 
\begin{equation}
n_g = \int_{0}^{\chi_s} n(z) \frac{dz}{d\chi^{\prime}} d\chi^{\prime}
\end{equation}
The fraction of the sky observed  in the weak lensing survey is assumed to be $f_{sky} = 0.5$ and we adopt  $\sigma_\epsilon = 0.4$ as the galaxy-intrinsic rms shear \cite{Hu_1999}. The factor $(2\ell +1)$ in the denominator counts the number of samples of  $C_{\kappa}^{\ell} $ for a given $\ell$.


On large scales the  redshifted \nh 21-cm signal from post reionization epoch known to be biased tracers of the underlying dark matter distribution. Assuming \nh  perturbations are generated by a Gaussian random process, and incorporating the effect of redshift space distortion the fluctuations of the 21-cm brightness temperature $\delta_T ({\bf r})$ in Fourier space may be written as

\[ \delta_T({\bf r}) = \frac{1}{(2\pi)^3} \int e^{i {\bf k }\cdot {\bf r}} \Delta_T({\bf k}) \] 
where  
\be 
\Delta_T( {\bf{k}} )  = \A (z) \left [ 1 + \beta_T \mu^2 \right ] \Delta ( {\bf k}) \ee

with \be
\A_{T} = 4.0 \, {\rm {mK}} \,
b_{T} \, {\bar{x}_{\rm HI}}(1 + z)^2\left ( \frac{\Omega_{b0}
  h^2}{0.02} \right ) \left ( \frac{0.7}{h} \right) \left (\frac{H_0}{H(z)} \right) 
\label{eq:21cmkernel}
\ee Here, $\Delta$ denotes the  fluctuations of the dark matter overdensity field in Fourier space, $b_T$ is a bias function and $ {\bar{x}_{\rm HI}}$ is the mean neutral fraction which is assumed to remain constant in the post-reionization epoch ($z<6$).We adopt the value $ {\bar{x}_{\rm HI}} = 2.45 \times 10^{-2}$ from
(\cite{Lanzetta_1995, P_roux_2003, Zafar_2013, Noterdaeme_2009}). 
We define a  quantity on  the sky 
\be {T} (\n) = \frac{1}{\chi_s - \chi_0}\sum_{\chi_0}^{\chi_s} \Delta \chi ~
\delta_{T}(\chi \n , \chi) \ee
as the integral of the 3D 21-cm brightness temperature field along the radial direction.

In a radio interferometric observation the quantity of interest is the complex
Visibilities which are Fourier transformation of the intensity distribution on
the sky. Using a "flat sky" approximation we define Visibilities as 
\be  V_{{T}}(\vu ) = \int d^2 \vtheta ~ a(\vtheta) {T} ( \vtheta ) ~e^{ -2\pi i \vu
  . \vtheta} ~~{\&}~~ V_{\kappa}( \vu ) = \int d^2 \vtheta ~ \kappa ( \vtheta )
~e^{ -2\pi i \vu . \vtheta} 
\ee 
where $ \vtheta$ is the angular coordinates on the flat sky plane,  $a(\vtheta)$ denotes the beam
function of the telescope measuring the angular coverage of the 21-cm survey. The mulipole $\ell$ is related to 
baseline $\vu$ as $ \ell = 2 \pi U$.
The aperture function $\widetilde{a}(\vu)$ is the Fourier transformation
of $ a(\vtheta)$.
Defining the cross-correlation angular power spectrum $ C^{T \kappa}(U)$ as
$\langle V_{{T}}( \vu )  V^{*}_{\kappa}( \vu' ) \rangle = C^{T \kappa}(U) $
we  have  \cite{Dash_2021} for a sharp aperture $\widetilde{a}(\vu)$, 
\be
\label{eq:crosssignal}
C^{T \kappa}(U) = \frac{1 }{\pi(\chi_s- \chi_0)} \sum_{\chi_o}^{\chi_s}  \frac{\Delta \chi}{\chi^2}    ~  \A_T \A_\kappa 
\int_0^{\infty} d\kpar   \left [ 1 + \beta_T  \frac{\kpar^2}{k^2} \right ]   P (k,\chi) 
\ee
with $k = \sqrt{\kpar^2 + \left (\frac{2 \pi \vu}{\chi} \right )^2 } $.

The auto-correlation angular power spectra  may be similarly written as \cite{Dash_2021}
\be
\label{eq:auto21}
C^{TT}  (U) = \frac{1 }{\pi(\chi_2- \chi_1)^2} \sum_{\chi_1}^{\chi_2}  \frac{\Delta \chi}{\chi^2}    ~  \A_T^2 
\int_0^{\infty} d\kpar   \left [ 1 + \beta_T  \frac{\kpar^2}{k^2} \right ] ^2  P (k,\chi) 
\ee
\be
\label{eq:autolensing}
C^{\kappa \kappa}  (U) = \frac{1 }{\pi} \int_{0}^{\chi_s}  \frac{d \chi}{\chi^2}    ~  \A_\kappa^2  
\int_0^{\infty} d\kpar    P (k,\chi) 
\ee

We follow the formalism in \cite{Dash_2021} and considered  the cross-correlation with the 21-cm signal averaged over the signals from redshift slices  to improve the signal to noise ratio.
As a note of caution we point out that working in the Fourier basis in the flat sky approximation necessarily makes the signal
nonergodic when we consider correlation between two time slices (due to time
evolution of all the relevant cosmological quantities). Further, one also notes the complications arising from the 
inseparability of the baseline ${\bf U}$ (transverse) from the frequency
(radial) in this formalism \cite{Sarkar_2017}.

The angular power spectrum for two redshifts separated by $\Delta z$ is known to
decorrelate very fast in the radial direction \cite{poreion7}.  In this work
we consider the summation in Eq: (\ref{eq:crosssignal}) over redshift
slices each of whose width is larger than the typical decorrelation length. Each term in the sum  can thus be thought of as an independent observation of the signal.  Thus the  noise in each term in the summation may be
thought of as an independent random variable and the mutual noise covaraiances
between the slices may be ignored.  Thus the errors in $C^{T \kappa}(U)$ is given by \be
\label{eq:variance}
\sigma_{_{T \kappa}}= \sqrt{ \frac {{(C^{\kappa \kappa} + \langle N^{\kappa } \rangle )(
    C^{TT} + \langle N^T \rangle )}}{   ({2\ell + 1})
  {N_c} }}  \ee
where $N_c$ is the
number of redshift slices over which tthe signal is averaged in Eq:
(\ref{eq:crosssignal}) and $ \langle N^{F_{HI}} \rangle $ and $\langle
N^{\kappa}\rangle $ corresponds to the average of the noise power spectrum for
$F_{HI}$ and $ \kappa$ respectively.

\label{eq:Fisher}

We compute the expected bounds on HS $f(R)$ gravity free parameter which measures the deviation from $\Lambda$CDM models. 
We have considered telescope specifications of the  upcoming SKA1-mid radio interfeoremeter. We have used the cosmological parameters from  Planck-2018 results  $(\Omega_{m{_0}},\Omega_{b_{0}},H_0,n_s,\sigma_8, \Omega_K) = (0.315,~0.0496,~67.4,~0.965,~0.811, ~0) $
from \cite{Planck2018} for our subsequent analysis.
The model galaxy  distribution function ($n(z,z_0)$) is adopted from \cite{takada2004cosmological, Huterer_2006}. The cross-correlation can only be computed in an overlapping volume for the weak lensing and 21-cm intensity mapping survey. 
We choose the frequency band $400-950$ $MHz$ of SKA1-mid since it corresponds to a redshift range that overlaps with  the redshift range of the weak lensing survey.  SKA1-mid has $250$ antennaes. The diameter of each antenna is taken to be $13.5$m  and system temperature ($T_{sys}$) assumed to be $40$ K for the entire redshift range. We also assume that full frequency band will be sub-divided into smaller frequency bands of 32 $MHz$. The details of the SKA1-mid telescope specifications including the baseline distribution can be found in the SKA website 
\footnote{https://www.skatelescope.org}.

\begin{figure}
\begin{center}
\includegraphics[width=8.5cm]{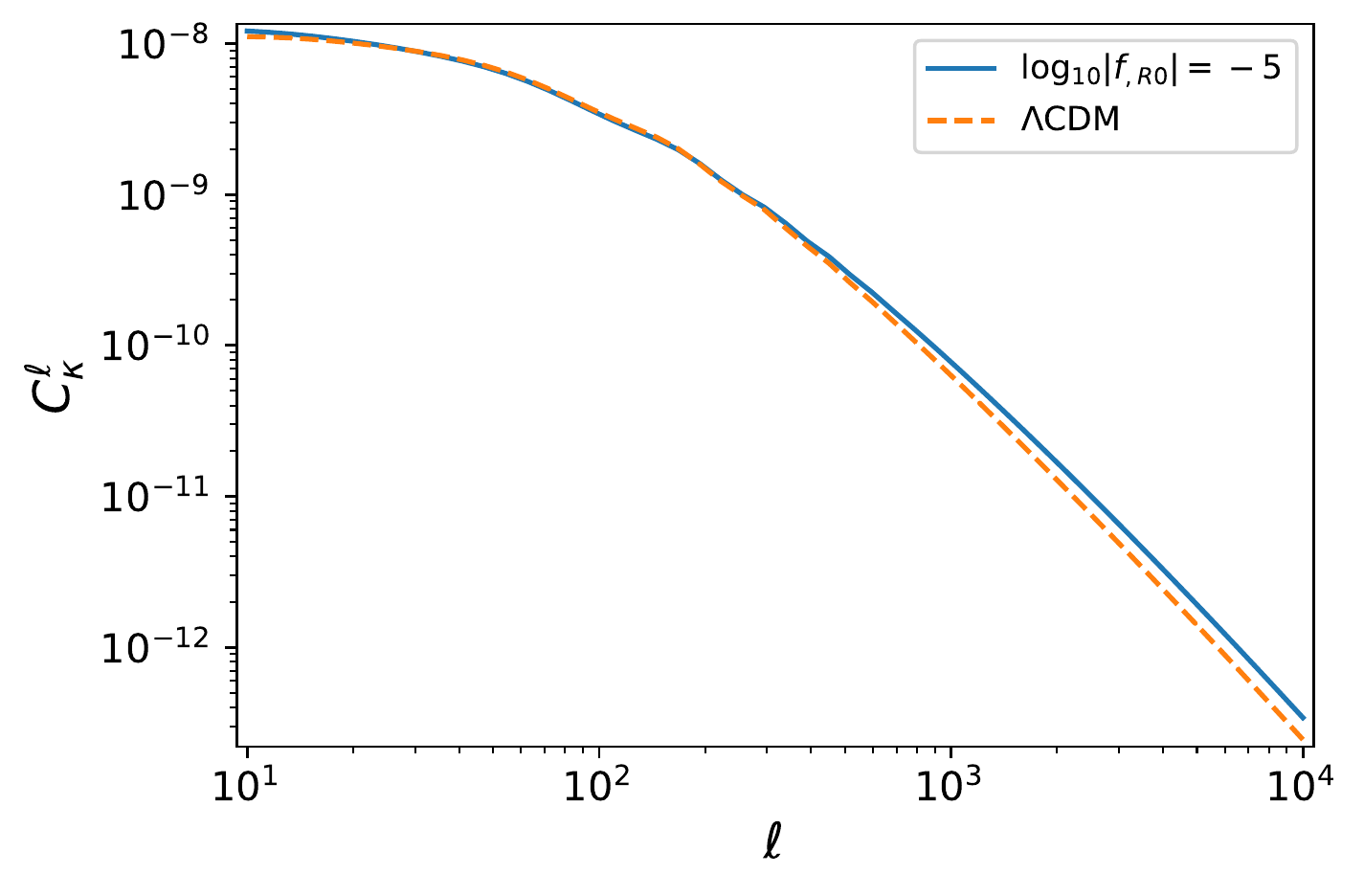}
\caption{The figure shows auto-correlation signal as a function of multiples for modified $f(R)$ model.The dotted line shows the $\Lambda$CDM prediction. The source redshift of a galaxy assumed to be $z_s =1.0$}
\label{fig:auto-correlation}
\end{center}
\end{figure}
\begin{figure}
\begin{center}
\includegraphics[width=7.5cm]{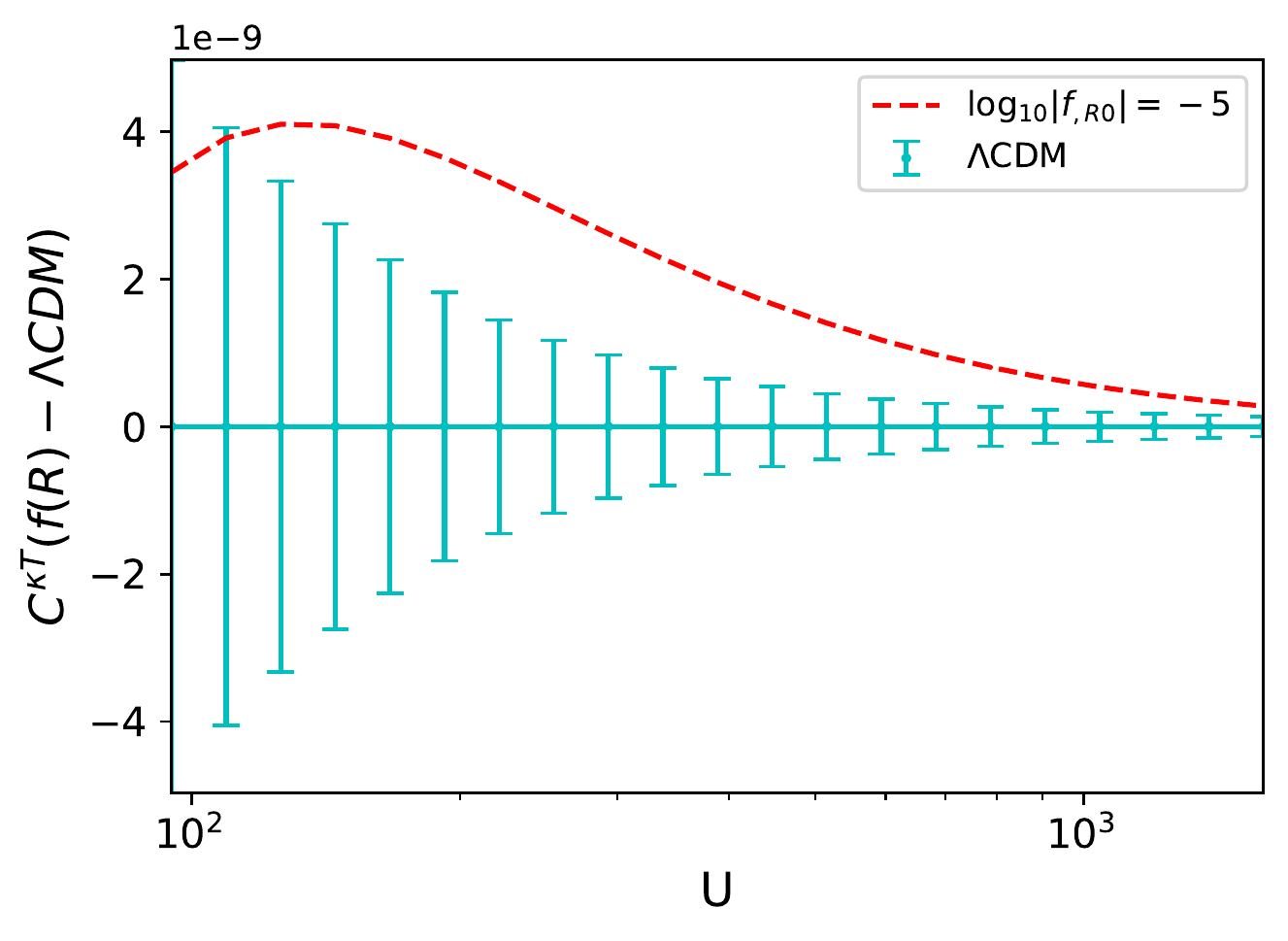}
\caption{The figure shows difference of \nh 21-cm - galaxy weak lensing cross-correlation power spectrum for the HS model with free parameter $\log_{10}|f_{,R0}|=-5$ from the standard $\Lambda$CDM. The 1-$\sigma$ error bars on $\Lambda$CDM shown assuming the galaxy density $n_g = 60arcmin^{-2}$ and observation time $T_{obs} = 600hrs$. }
\label{fig:cross-correlation}
\end{center}
\end{figure}

Cross-correlation of CMBR weak lensing and \nh 21-cm power spectrum has been studied  earlier \cite{GSarkar_2010, Dash_2021}. In this paper we address the cross-correlation with galaxy weak lensing. A typical galaxy weak lensing survey is different from CMBR weak lensing survey for the following reason. The CMBR temperatures are drawn from a Gaussian distribution, where the galaxies are the tracers of the underlying matter distribution, which at least small scales completely non-linear. However we have not incorporated the effects of non-linearity in our analysis as  we are working in the regime of the linear perturbation theory. Secondly,  the galaxy surveys are purely 3D while CMB anisotropies are in general a function of angular position $\ell$ on sky. 
The Figure:(\ref{fig:auto-correlation}) shows theoretically expected convergence auto-correlation angular power spectrum signal for $\Lambda$CDM and HS model with free parameter $\log_{10}|f_{,R0}|=-5$ for reference. The source redshift of galaxy assumed to be $z_s=1$. It can be seen that on larger scales the $f(R)$ model predictions agree with $\Lambda$CDM. A significant deviation from classical GR is only found beyond a scale $\ell>200$ because of the scale dependent growing mode. Similar results are obtained from simulations in \cite{higuchi2016imprint,li2018galaxy}.  We also note that the deviation from $\Lambda$CDM in a range of scale $200<\ell<3000$ is  typically around $10-15\%$.

We are interested in the cross-correlation signal of \nh 21-cm and galaxy lensing. The cross-correlation signal takes the same shape as of convergence auto-correlation signal. We have computed the cross-correlation signal using the equation (\ref{eq:crosssignal}).  The Figure (\ref{fig:cross-correlation}) shows the difference of the \nh 21-cm and galaxy weak lensing angular cross-correlation power spectrum for HS parametrization with $\log_{10}|f_{,R0}|=-5$ from $\Lambda$CDM. The 1-$\sigma$ error bars on $\Lambda$CDM shows the HS model with $\log_{10}|f_{,R0}|=-5$ can be differentiate from $\Lambda$CDM at a level of $>2\sigma$ sensitivity using galaxy density $n_g=60 {\rm arcmin}^{-2}$ and radio interferometric observation time $T_{obs}=600hrs$.

\begin{table}[h]
\label{table111}
\begin{center}
\caption{The the $68 \%$ ($ 1 -\sigma$) Constraints of $\log_{10}|f_{,R0}|$ and $\Omega_{m0}$ from \nh 21-cm and galaxy lensing cross power spectrum}
\begin{tabular}{c|c c} 
 \hline 
 \hline 
  Model &  $ \log_{10}| f_{R0} |$  & $ \Omega_{m0} $  \\ 
 \hline
 HS-$f(R)$ & $-5\pm 0.59$  & $0.315 \pm 0.10$ \\
 \hline
  \hline
    \end{tabular}
    \end{center}
    	\end{table}
The Fisher analysis is used to put bound on the parameter $\log_{10}|f_{,R0}|$ using the cross-correlation signal. Assuming the fiducial value of $\log_{10}|f_{,R0}|=-5$ and marginalizing over the overall amplitude, redshift distortion parameter ($\beta_T$) we found the the $1-\sigma$ bounds on $\log_{10}|f_{,R0}|$ as shown in the above table.

\section{Crosscorrelation of 21-cm signal with Lyman-$\alpha$ forest}

\subsection*{Lyman-$\alpha$ forest power spectrum}
Lyman-$\alpha$ forest traces out the small fluctuations in the \nh density in the IGM along the line of sight (LoS) to distant background quasars and shows an absorption features in the quasar spectra. The quantity of interest is the transmitted QSO flux through the Lyman-$\alpha$. The fluctuating Gunn-Peterson effect allows us to write
\be
\F = \bar{\F} e^{-A(1+\delta)^\Gamma}
\ee
where $\bar{\F}$ denotes the mean transmitted flux, $\Gamma$ depend on the slope of the temperature-density power spectrum and the factor $A$ depends on the \nh photoionization rate, which is difficult to measure independently and assumed to be nearly $\sim 1$. Several simulation works of Lyman-$\alpha$ forest shows the transmitted flux $\delta_\F = (\bar{\F}-F)/\bar{F}$ $\propto$ $\delta$ \cite{Carucci_2017}. 

The influence of $f(R)$ gravity theory in the Lyman-$\alpha$ forest power spectrum has been studied extensively \cite{arnold2015lyman, brax2019lyman}. Fitting formulas
for Lyman-$\alpha$ forest power spectrum $(P_{\F\F}(k))$ are usually written in terms of the matter power spectrum $P(k)$ with several prefactors to match numerical simulations. We follow \cite{brax2019lyman} to model the Lyman-$\alpha$ power spectrum in $f(R)$ gravity theory. The Lyman-$\alpha$ power spectrum can be written in terms of matter power spectrum as follows
\be
\label{eqn:lymanps}
P_{\F\F}(k,z) = \frac{ (1+\beta_{_{\F}} \mu^2)^2}{  (1+f_g k_{\parallel} /k_{NL}) }  P(k,z)  e^{-(k_{\parallel}  / k_{th})^2}
\ee
where $\mu$ is the cosine of the angle between LoS $(\hat{n})$ and the wave vector $(\vec{k})$ so that  $\mu = \hat{k}.\hat{n}= k_{\parallel}/k$. 
Here $\beta_{_{\F}}$ is the large scale aniosotropy parameter or so called the redshift distortion factor and $k_{th}$ is the thermal brodening cutoff wave number. We will use Eq : \ref{eqn:lymanps} to compute the 3D and 1D Lyman-$\alpha$ auto correlation power spectrum.
The Eq : \ref{eqn:lymanps} gives the 3D Lyman-$\alpha$ power spectrum in the redshift space. The observed 1D power spectrum along LoS is given by the standard integral 
\be
\label{eq:1dLymanps}
P^{1D}_{\F\F} (k_{\parallel}) = \frac{1}{(2\pi)^2} \int dk_{\perp} P_{\F\F}(k)
\ee

\begin{figure}[h]
\begin{center}
\includegraphics[width=8.5cm]{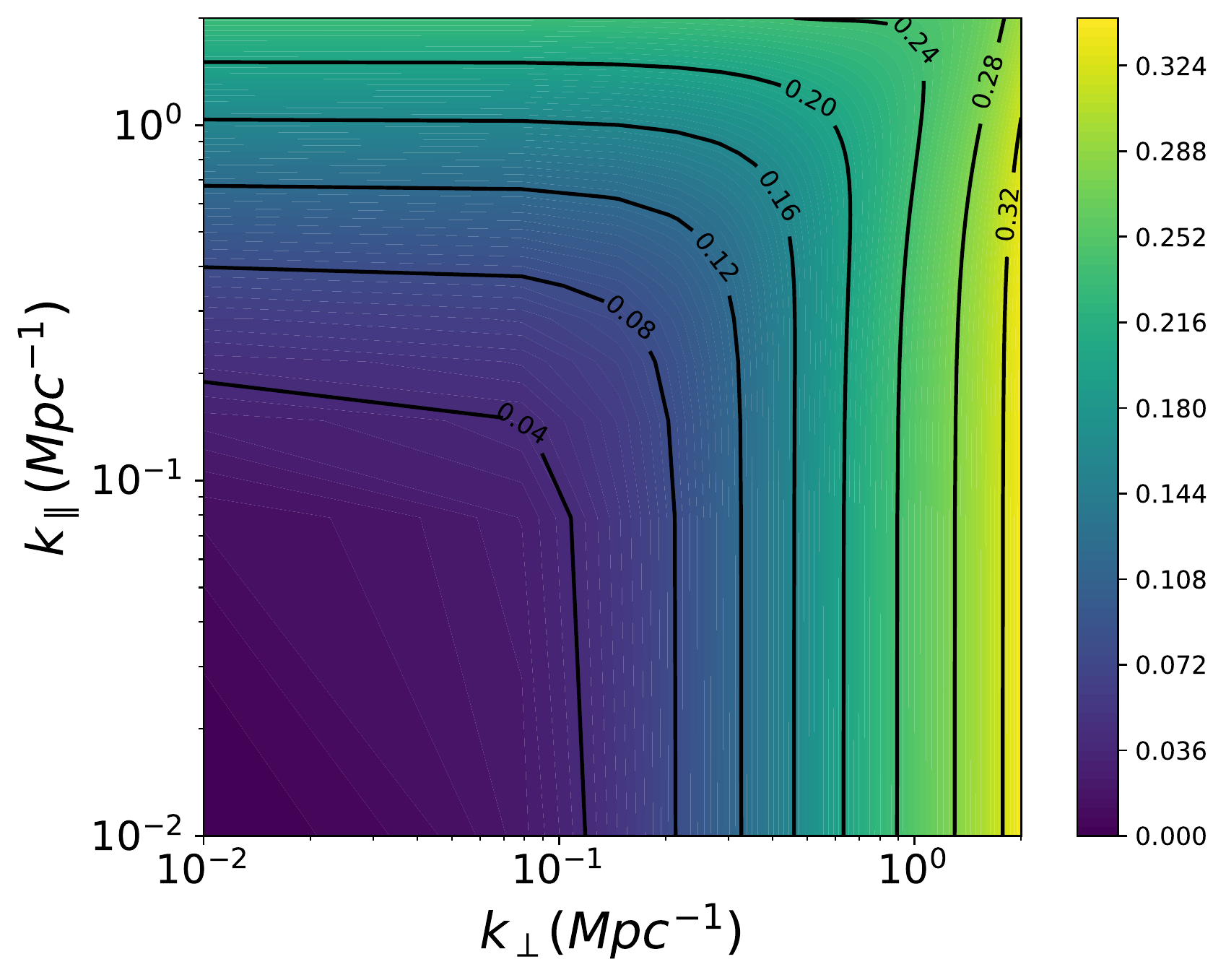}
\caption{Figure shows the 3D cross-correlation power spectrum power spectrum in redshift space for $f(R)$ gravity at a fiducial redshift $z = 2.3$. The asymmetry in the signal is indicative of redshift space distortion.}
\label{fig:crosssignal}
\end{center}
\end{figure}

Both Lyman-$\alpha$ and the \nh 21-cm signal from the post reionization epoch are extremely useful tools to probe underlying theory of gravity and put stringent constraints on cosmological parameters individually. However on large scale both trace the dark matter density field motivating us to investigate their cross-correlation signal \cite{Guha_Sarkar_2010}. The  cross-correlation of the Lyman-$\alpha$ and \nh 21-cm signal has been studied for the $\Lambda$CDM model extensively \cite{Guha_Sarkar_2010, Sarkar_2015, Carucci_2017, Sarkar_2018, Sarkar_2019}. In this paper we shall extend  it to $f(R)$ gravity models. 
The Lyman-$\alpha$ and \nh 21-cm signal can be written using the formalism in \cite{Sarkar_2015} and equation (\ref{eqn:lymanps}).
 We choose  a fiducial redshift $z=2.3$ for this analysis. Figure (\ref{fig:crosssignal}) shows the 3D cross correlation power spectrum in $(k_{\perp},k_{\parallel})$ plane for $\log_{10}|f_{,R0}| = -5$. The asymmetry in cross signal arises because of Kaiser effect in the redshift space.  However the deviation of asymmetry is much enhanced than the auto correlation signal. 
The Figure (\ref{fig:crosssignal}) shows the 3D Lyman-$\alpha$ and \nh 21-cm  cross correlation power spectrum at a fiducial redshift $z =2.3$. The fiducial redshift chosen to be $z=2.3$ as the QSO distribution is known to peaks at $z=2.25$ and falls off as we move away from peak  \cite{abolfathi2018fourteenth}. The deviation of spherical symmetry in power spectrum arises because of the linear redshift space distortion parameter $\beta_{\F}$ and $\beta_{T}$.
We next use the cross correlation signal to put constraints on the parameter $\beta_{T}(k,z)$.
We have used the cosmological parameters from  Planck-2018 results $(\Omega_{m{_0}},\Omega_{b_{0}},H_0,n_s,\sigma_8, \Omega_K) = (0.315,~0.0496,~67.4,~0.965,~0.811, ~0) $
from \cite{Planck2018} for our subsequent analysis.
We consider a radio interferometric array  for the 21-cm observations mimicking the SKA1-mid. The SKA1-mid is one of the three different instruments that will be built as a part of the SKA telescope. SKA1-mid has $250$ antennaes. The diameter of each antenna is taken to be $13.5$m  and system temperature ($T_{sys}$) assumed to be $40$ K for the  redshift $z=2.3$. We also assume that full frequency band will be sub-divided into smaller frequency bands of 32 MHz. 
For Lyman-$\alpha$ forest observation we have used the quasar number of distribution from DR14 of SDSS \cite{abolfathi2018fourteenth}. It has a total angular coverage of 14,555 deg$^2$ and we assumed the of QSO number density $\bar{n}=60deg^{-2}$. 
Each spectra is assumed to have been measured at $>3 \sigma$. sensitivity.
\begin{figure}[h]
\begin{center}
\includegraphics[width=8.5cm]{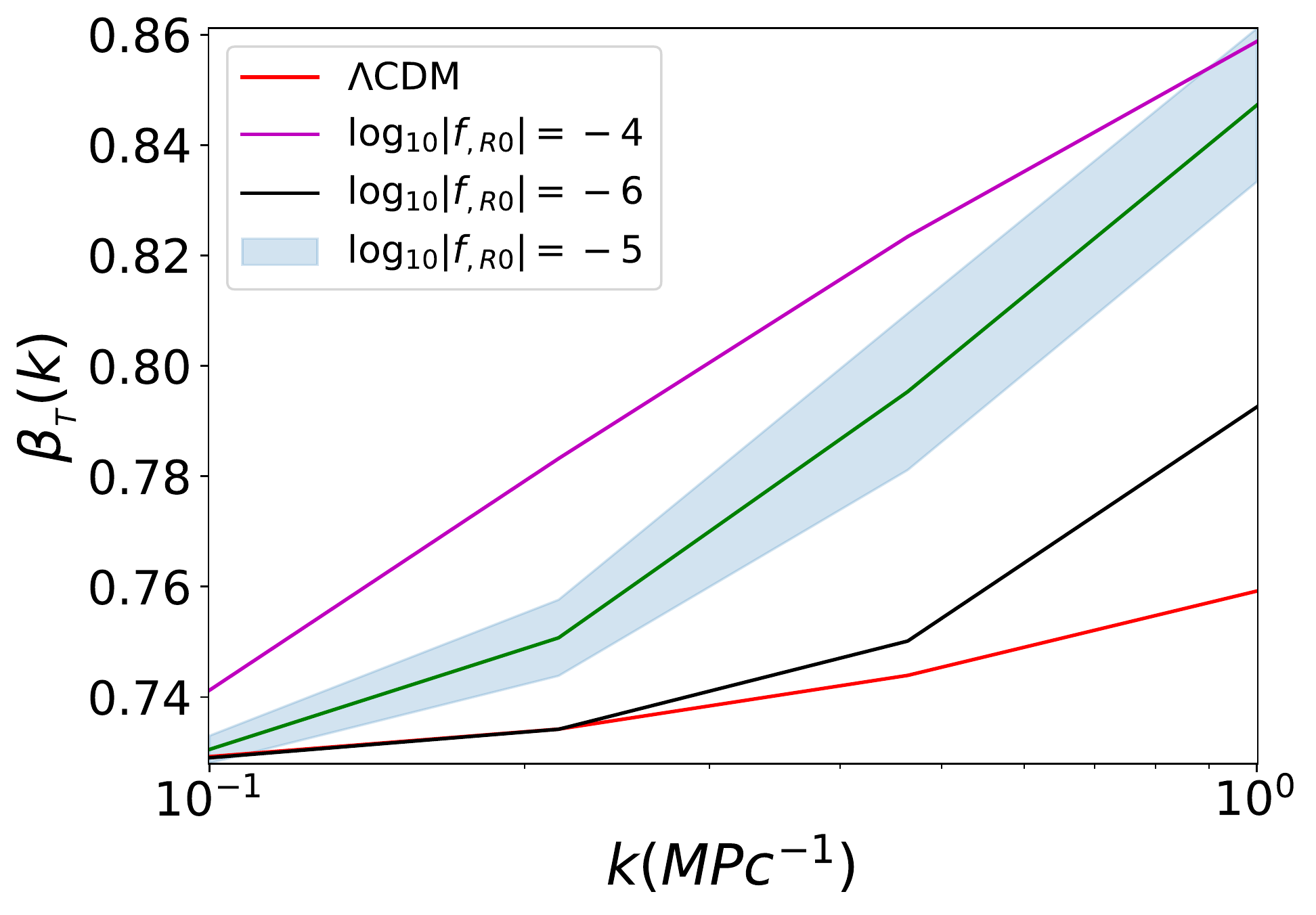}
\caption{The figure shows the redshift distortion parameter $\beta_{T}$ at a fiducial redshift $z = 2.3$.The $1-\sigma$ marginalized error is shown by the shaded region on the top of fiducial $f(R)$ gravity $\log_{10} |f_{,R 0}| = - 5$ model. We also shown the different parameterized $f(R)$ gravity models and $\Lambda$CDM for comparison.}
\label{fig:constraints}
\end{center}
\end{figure}

We have divided the  $k-$range from $0.1<k<1$ into $4$ $k-$bins. We perform Fisher matrix analysis for the following  parameters -the binned values of $\beta_T$, the  overall normalization factor $(\bar{\N})$, distortion factor ($\beta_{_{\F}})$, third order component of polynomial bias $b_{T}$.  We have marginalized over all the parameters except the four values of $\beta_T$.

Fig:(\ref{fig:constraints}) shows the $\beta_{T}(k,z)$ for $f(R)$ gravity models. The shaded region corresponds to the $1-\sigma$ error projection for the fiducial $\log_{10} |f_{,R 0}| = - 5$ gravity model. At large scale all $f(R)$ gravity theories matches with standard $\Lambda$CDM model. However we find that on small scales beyond $(k>0.5)$, the $\log_{10} |f_{,R 0}| = - 5$ model can be distinguished from $\Lambda$CDM model at a level of $3-\sigma$ sensitivity if we consider $2$ $k-$bins  for $500 \times 60$hrs observation with 60 independent pointings. But other $f(R)$ gravity models are not very much distinguishable $(<3-\sigma)$ throughout the $k$ range. This is because where at very higher redshifts we expect all the modified gravity theories matches to our standard concordance $\Lambda$CDM model and deviation from it is much  smaller.
Instead of constraining the binned function $\beta_{T}(k)$, we investigate the possibility of putting bounds on $\log_{10} |f_{,R0}|$ from the given observation. Marginalizing over the overall amplitude of the power spectrum, we are thus interested in two parameters $(\Omega_{m0}, \log_{10} |f_{,R0}|)$.  The error projections given below.
\begin{table} [h]
\begin{center}
\caption{The $68 \%$ ($1-\sigma $) marginalized errors on $\log_{10}| f_{,R0} |$  and $ \Omega_{m0} $ from the 21-cm and Lyman-$\alpha$ cross-correlation.}
\begin{tabular}{c|c| c} 
 \hline 
 \hline 
   Model &  $ \log_{10}| f_{R0} |$  & $ \Omega_{m0} $  \\ 
 \hline
 $f(R)$ & $ ~~~-5 \pm 0.29~~$  & $~~~  0.315 \pm 0.012~~~ $ \\
 \hline
\end{tabular}
\end{center}
\end{table}

\section{Conclusion}
Einstein's relativity has been extremely well tested on solar system scales \cite{chiba2007solar,shapiro2004measurement,bertotti2003test}. 
The $f(R)$ modification often confronts the strong agreement of general relativity on such small scales. Einstein's relativity can be recovered and solar system tests be evaded by the chameleon mechanism \cite{gu2011solar,Khoury_2004,capozziello2008solar}.
Effectively, this implies that $f(R)$ differs very little from $R$ on solar system scales. It has been shown that Hu-Sawicki $f(R)$ gravity models agree well with the late time cosmic acceleration without invoking a cosmological constant and satisfies both cosmological and solar-system tests in the weak field limit \cite{de2010f}. However, solar-system tests alone put only weak bounds on these models \cite{Hu_2007} and there is a great variability of model parameters. We have shown that the 21 cm intensity mapping instruments like SKA1 will be capable of constraining the a field value $\log_{10}|f_{,R0}| = -5 \pm 0.62$ of $68\%$ confidence. This is an order of magnitude tighter than constraints currently available from galaxy cluster abundance \cite{ferraro2011cluster}. Further \cite{Cataneo_2015} showed that marginalized $95.4\%$ the the upper limit on $\log_{10}|f_{,R0}|  = -4.79$ using the Cluster+Planck+WMAP+Lensing+ACT+SPT+SN+BAO data. Joint analysis of 21-cm intensity mapping with the above observation probe shall be able to narrow down the current constraints.
We note that the low redshift departure of  $f(R)$ gravity from GR  predictions is small and better modeling is needed to invoke non-linear chameleon suppression for tighter constraints on $f(R)$ models.

The radio-interferometric observation of the post-reionization \nh 21-cm signal, thus holds the potential of providing  robust constraints on $f(R)$ models.
We have seen that the error projections from both the auto-correlation and cross-correlation signals provide competitive bounds on $f(R)$ models.
Several observational aspects, however, plague the detection of the 21-cm signal. 
We have evaded the key observational challenge arising from large astrophysical foregrounds that plague the signal.
 Astrophysical foregrounds from both galactic and extra galactic sources plague the signal and significant amount of foreground subtraction is
required before one may detect the signal. Several methods of subtracting foregrounds have been suggested (see \cite{2011MNRAS.418.2584G} and citations in this work)
Cross-correlation of the 21-cm signal has also been proposed as a way to mitigate the issue of large foregrounds \cite{Guha_Sarkar_2010, Sarkar_2015}.
The cosmological origin of the 21-cm signal may only be ascertained in a cross-correlation.  
The foregrounds appear as noise in the cross-correlation and may be tackled by considering larger survey volumes. 
Further, man made radio frequency interferences (RFIs), calibration errors and other observational  systematics  inhibits the sensitive detection of the \nh 21-cm signal. A detailed
study of these observational aspects shall be studied in a future work.
We conclude by noting  that future observation of the redshifted \nh 21-cm signal shall be an important addition to the different cosmological probes aimed towards measuring possible modifications to Einstein's gravity. This shall  enhance our understanding of late time cosmological evolution and structure formation.

\bibliographystyle{apsrev4-1}
\bibliography{references}

\end{document}